\documentclass[11pt]{article}

\pdfoutput=1
\usepackage{epsfig,amssymb,bm, graphicx, amsmath}
\usepackage{epstopdf} 

\usepackage[normalem]{ulem}
\usepackage{jcappub}
\usepackage{color}

  \def\be{\begin{equation}}
  \def\ee{\end{equation}}    
  \def\ba{\begin{eqnarray}}
  \def\ea{\end{eqnarray}}

  \def\M{\mathcal{M}}
  \def\Mh{\mathcal{M}^{\rm h}}
  \def\Mf{\mathcal{M}^{\rm f}}
  
  \newcommand*{\Msun}{\ensuremath{\, M_{\odot}}}

  \newcommand{\Expect}[1]{\left\langle #1 \right\rangle}

  \newcommand*{\mysub}[2]{\ensuremath{#1_{\mathrm{#2}}}}
  \newcommand*{\Omegam}{\mysub{\Omega}{m}}
  \newcommand*{\Omegab}{\mysub{\Omega}{b}}

  \newcommand*{\secref}{Section~}
  \newcommand*{\figref}{Figure~}

  \title{X-ray Cluster Constraints on Non-Gaussianity}
  \author[a,f]{Sarah Shandera,}
  \author[b,f]{Adam Mantz,}
  \author[c]{David Rapetti,}
  \author[d,e]{Steven W. Allen}
  \affiliation[a]{Institute for Gravitation and the Cosmos, The Pennsylvania State University, University Park, PA 16802, USA}
  \affiliation[b]{Kavli Institute for Cosmological Physics, University of Chicago, 5640 South Ellis Avenue, Chicago, IL 60637, USA}
  \affiliation[c]{Dark Cosmology Centre, Niels Bohr Institute, University of Copenhagen, Juliane Maries Vej 30, 2100 Copenhagen, Denmark}
  \affiliation[d]{Kavli Institute for Particle Astrophysics and Cosmology, Stanford University, 452 Lomita Mall, Stanford, CA 94305, USA}
  \affiliation[e]{SLAC National Accelerator Laboratory, 2575 Sand Hill Road, Menlo Park, CA 94025, USA}
  \affiliation[f]{\textcolor{blue}{Corresponding authors}}
  \emailAdd{shandera@gravity.psu.edu, amantz@kicp.uchicago.edu, drapetti@dark-cosmology.dk, swa@stanford.edu}
  \abstract{We report constraints on primordial non-Gaussianity from the abundance of X-ray detected clusters. Our analytic prescription for adding non-Gaussianity to the cluster mass function takes into account moments beyond the skewness, and we demonstrate that those moments should not be ignored in most analyses of cluster data. We constrain the amplitude of the skewness for two scenarios that have different overall levels of non-Gaussianity, characterized by how amplitudes of higher cumulants scale with the skewness. We find that current data can constrain these one-parameter non-Gaussian models at a useful level, but are not sensitive to adding further details of the corresponding inflation scenarios. Combining cluster data with Cosmic Microwave Background constraints on the cosmology and power spectrum amplitude, we find the dimensionless skewness to be $10^3\mathcal{M}_3=-1^{+24}_{-28}$ for one of our scaling scenarios, and $10^3\mathcal{M}_3=-4\pm7$ for the other. These are the first constraints on non-Gaussianity from Large Scale Structure that can be usefully applied to any model of primordial non-Gaussianity. The former constraint, when applied to the standard local ansatz (where the $n$-th cumulant scales as $\mathcal{M}_n\propto\mathcal{M}_3^{n-2}$), corresponds to $f^{\rm local}_{\rm NL}=-3^{+78}_{-91}$. When applied to a model with a local-shape bispectrum but higher cumulants that scale as $\mathcal{M}_n\propto\mathcal{M}_3^{n/3}$ (the second scaling scenario), the amplitude of the local-shape bispectrum is constrained to be $f^{\rm local*}_{\rm NL}=-14^{+22}_{-21}$. For this second scaling (which occurs in various well-motivated models of inflation), we also obtain strong constraints on the equilateral and orthogonal shapes of the bispectrum, $f_{NL}^{\rm equil} = -52^{+85}_{-79}$ and $f_{NL}^{\rm orth} = 63^{+97}_{-104}$. This sensitivity implies that cluster counts could be used to distinguish qualitatively different models for the primordial fluctuations that have identical bispectra.
  }
  \preprint{IGC-13/04-1}

\begin{document}
  \maketitle

  \section{Introduction}
  The Large Scale Structure of the late Universe depends on a rich array of physics: the spectrum of primordial curvature inhomogeneities, the cosmological evolution of the Universe, the rules governing the growth of structure, and particle physics, chemistry and thermodynamics within individual stars, galaxies, and galaxy clusters. Extracting details of the primordial fluctuations is necessarily a difficult problem, but fortunately there are several complementary observables available to us. In this work, we use measurements of the mass and redshift distribution (the mass function) of a sample of galaxy clusters to constrain primordial non-Gaussianity. We demonstrate that this is a complementary probe to the Cosmic Microwave Background (CMB) bispectrum and the halo bias because it is sensitive to different aspects of the primordial non-Gaussianity.

  Most work on cosmological constraints from clusters has focused on dark energy \cite{Henry:2008cg,Vikhlinin:2008ym,Mantz:2009fw}. Here we apply the substantial progress made on characterizing the mass--observable relations in that context to study primordial non-Gaussianity. We use 237 X-ray bright clusters detected in the ROSAT All-Sky Survey \cite{Truemper:1993} and the analysis techniques of Mantz et al. \cite{Mantz:2009fx} to investigate two one-parameter models for the non-Gaussian curvature perturbations and four two-parameter models. The clusters in this sample have redshifts up to $z=0.5$ and masses of order $10^{14}$--$10^{15}\Msun$. With the semi-analytic, non-Gaussian mass function extended to include terms beyond the skewness, we find constraints that are completely consistent with Gaussian statistics for the primordial fluctuations. However, contrary to some of the expectations in the literature \cite{Cunha:2010zz, Pillepich:2011zz,Mak:2012yb}, we find error bars small enough to indicate that cluster counts can provide complementary information to current constraints from the CMB bispectrum \cite{Bennett:2012fp, Ade:2013tta} and from the galaxy bias \cite{Slosar:2008hx,Xia:2010pe,Xia:2011hj,Giannantonio:2013uqa}. Our results are consistent with but tighter than those recently obtained using clusters detected by the South Pole Telescope (SPT) \cite{Benson:2011uta, Williamson:2011jz} and clusters selected from the Sloan Digital Sky Survey (SDSS) \cite{Mana:2013qba}.

  Statistics of the primordial fluctuations beyond the homogeneous and isotropic power spectrum are an extremely important source of information about the very early Universe -- so much so that pursuing limits on non-Gaussianity down to the minimal levels expected from single field slow-roll inflation is an important task. Galaxy clusters form from rare primordial over-densities, with more massive and higher redshift clusters tracing rarer initial fluctuations, and so their abundance is sensitive to non-Gaussianity in the primordial inhomogeneities. Constraints from cluster number counts are complementary to other probes of non-Gaussianity in three ways: they probe smaller scales than the CMB or galaxy bias do (cluster constraints are at $k\approx0.1$--$0.5\, h/{\rm Mpc}$), they are sensitive to {\it any} non-Gaussianity (including any shape for the primordial bispectrum), and they are sensitive to higher order cumulants of the probability distribution of the primordial inhomogeneities. 

  The non-Gaussian statistics of the primordial perturbations are completely described by the set of correlation functions $\langle\Phi(\vec{k}_1)\Phi(\vec{k}_2)\dots\Phi(\vec{k}_n)\rangle_c$, where $\Phi(k)$ is the Bardeen potential in momentum space, and $n$ runs from 3 to infinity. The subscript $c$ stands for `connected' and picks out the parts of the correlations that vanish for a Gaussian field. Clearly, a single parameter cannot describe this series of functions. However, if the field is weakly non-Gaussian, the three-point function is likely to generate the strongest observational signal, so many non-Gaussian models are classified by naming the configuration of the three-point function and its amplitude. The Wilkinson Microwave Ansiotropy Prove (WMAP) satellite, for example, reports constraints on the parameters $f_{\rm NL}^{\rm local}$, $f_{\rm NL}^{\rm equil}$, $f_{\rm NL}^{\rm orth}$, where the labels `local', `equilateral', and `orthogonal' refer to specific, scale-independent functional forms assumed for the three-point correlation. Translation invariance requires that the three-point correlation has a factor $\delta_D^3(\vec{k}_1+\vec{k}_2+\vec{k}_3)$ (a Dirac delta function), similar to the term $\delta_D^3(\vec{k}_1+\vec{k}_2)$ in the power spectrum. The three-point correlation is then just a function of two independent momenta and is called the bispectrum.

  The local, equilateral and orthogonal bispectra are shown in Eq.(\ref{templates}) below. Interestingly, though, object number counts are not sensitive to the details of the bispectrum's momentum dependence. Instead, only the integrated moments of the smoothed density fluctuations $\delta_R$ are relevant. For example,
  \be
  \langle\delta^3_R\rangle=\int\frac{d^3{\vec{k}_{1}}}{(2\pi)^3}\int\frac{d^3{\vec{k}_{2}}}{(2\pi)^3}\int\frac{d^3{\vec{k}_{3}}}{(2\pi)^3}M(k_1,R,z)M(k_2,R,z)M(k_3,R,z)\langle\Phi(\vec{k}_1)\Phi(\vec{k}_2)\Phi(\vec{k}_3)\rangle_c
  \label{eq:d3R}
  \ee
  where the terms $M(k_i,R,z)$ contain a window function, the corresponding factors from the Poisson equation, the transfer function and the growth factor converting the linear perturbation in the gravitational potential to the smoothed density perturbation. (The full expressions for these quantities can be found in Appendix \ref{appendix:MF}.) We characterize the non-Gaussianity by the dimensionless ratios of the cumulants of the density field
  \be
  \label{eq:Mns}
  \mathcal{M}_{n,R}=\frac{\langle\delta_R^n\rangle_c}{\langle\delta_R^2\rangle^{n/2}}
  \ee
  which are by construction redshift independent and nearly independent of the smoothing scale, $R$, if the primordial bispectrum is scale independent.\footnote{Scale independence means that the bispectrum (e.g., those in Eq.(\ref{templates})) contains no length scale other than the factors $k_i^{-1}$ in the $P(k_i)$ terms.}

  When cumulants beyond the skewness (correlations beyond the bispectrum) are relevant, a one-parameter model is only useful if we can use it to specify the amplitude of {\it all} the correlations. In this paper we use $\mathcal{M}_3$ and a choice for how higher moments scale with $\mathcal{M}_3$ to describe non-Gaussian fluctuations. The scalings we consider are motivated by particle physics models of inflation, and our constraints on the total dimensionless skewness can always be re-written in terms of a particular bispectrum using Eq.(\ref{eq:d3R}) and Eq.(\ref{eq:Mns}).

  Most previous work on the utility of cluster counts to constrain non-Gaussianity has focused on the local ansatz \cite{Salopek:1990jq, Komatsu:2001rj}, where one assumes that the non-Gaussian field $\Phi(x)$ is a simple, local transformation of a Gaussian field $\Phi_\mathrm{G}(x)$:
  \be
  \label{eq:localAnsatz}
  \Phi(x)=\Phi_\mathrm{G}(x)+f_{\rm NL}^{\rm local}[\Phi_\mathrm{G}(x)^2-\langle\Phi_\mathrm{G}(x)^2\rangle].
  \ee
  In this useful model, $f_{\rm NL}^{\rm local}$ is the single parameter that all correlation functions depend on, and the cumulants scale as $(f_{\rm NL}^{\rm local})^{n-2}$. Non-Gaussianity of the local type has a bispectrum that most strongly correlates Fourier modes of very different wavelengths. This particular mode coupling generates strong signals in other large scale structure observables -- most notably introducing a scale dependence in the bias of dark matter halos, luminous red galaxies and quasars \cite{Dalal:2007cu,Slosar:2008hx,Giannantonio:2013uqa}. For this reason, papers that have analyzed the potential for future surveys to constrain non-Gaussianity have largely focused on non-Gaussianity captured by the local ansatz only, and on the superior constraints from the bias compared to number counts. The bias may allow us to probe $\Delta f_{\rm NL}^{\rm local}\sim\mathcal{O}(1)$ in the near future \cite{Fedeli:2010ud,Giannantonio:2011ya}, although this optimism is still subject to a full understanding of the relevant systematics \cite{Pullen:2012rd, Huterer:2012zs, Giannantonio:2013uqa}. Regardless, the motivation for looking at cluster number counts to constrain a scale-invariant, local ansatz is certainly weak. However, single field models of inflation cannot produce large local non-Gaussianity over a wide range of scales \cite{Creminelli:2004yq}. Since non-Gaussianity that does not strongly correlate modes of very different wavelength is {\it not} particularly detectable in the galaxy bias \cite{Wagner:2011wx,Xia:2011hj, Agarwal:2013}, that measure alone tests only a subset of viable inflation models. Furthermore, number count constraints are sensitive to higher order correlation functions. As we will demonstrate here, number counts can distinguish between scenarios with indistinguishable bispectra that nonetheless are generated by qualitatively different inflationary physics. That is, a model may have a local-shape bispectrum but higher order moments that do not scale as those from the standard local ansatz do (e.g., \cite{Barnaby:2012tk}). In this case, number counts or other measurements sensitive to higher moments will provide complementary information to the halo bias constraints. For the rest of the paper, we will assume that $f_{\rm NL}^{\rm local}$ is a parameter measured from the bispectrum alone (see Eq.(\ref{templates}) below), which does not imply the entire series of correlations that Eq.(\ref{eq:localAnsatz}) generates, unless we specify the local ansatz as our model. This use of $f_{\rm NL}^{\rm local}$  is more in line with how it is observationally defined (e.g., in analyses of the CMB and halo bias). 

  This paper provides the first constraint on primordial non-Gaussianity from X-ray detected clusters, and the first large scale structure constraint that can be usefully applied to any model for primordial non-Gaussianity. In the next Section, we define our parametrization of the effects of primordial non-Gaussianity on cluster abundance, using a semi-analytic non-Gaussian mass function in terms of a single new parameter. In Section \ref{sec:2param} we discuss several theoretically motivated extensions to two parameters. We present our results for the one and two parameter models in Section \ref{sec:results}. In Section \ref{sec:OtherWork}, we discuss how our results compare to previous results and forecasts in the literature, including the SPT constraints reported from two small samples of Sunyaev-Zel'dovich (SZ) detected clusters \cite{Benson:2011uta, Williamson:2011jz} and those from a large sample of optically selected clusters \cite{Mana:2013qba}, the SDSS maxBCG cluster catalogue \cite{Koester:2007}. We summarize in Section \ref{sec:Conclude}. Appendix \ref{appendix:MF} contains the details of our semi-analytic mass function prescriptions as well as others that exist in the literature.\footnote{We have shared some computer code helpful for evaluating our mass function; see Appendix~\ref{appendix:MF}.} Quoted error intervals are always 68.3\% confidence, unless otherwise specified. 

  \section{The effect of Primordial non-Gaussianity on object number counts}
   \label{sec:mfcns}
  Our basic tool is a series expansion for the ratio of the non-Gaussian mass function to the Gaussian one. The expansion we use is based on a Press-Schechter model for halo formation applied to non-Gaussian probability distributions for the primordial fluctuations. A detailed derivation of the non-Gaussian mass function we use is given in Appendix \ref{appendix:MF} and was developed in \cite{LoVerde:2007ri, Barnaby:2011pe,Shandera:2012ke}. The weakly non-Gaussian probability distributions that the mass function is based on are asymptotic expansions that deviate substantially from the actual probability density function (PDF) for sufficiently rare fluctuations. Fortunately, our cosmology is already sufficiently constrained to determine that the clusters in our sample do not lie in that regime. However, the clusters {\it are} sufficiently rare that truncating the expansion below at a single term (the skewness) is not sufficient to test the full range of models that are only as skewed as current CMB constraints allow. 

  We add non-Gaussianity to the cosmology by considering a mass function of the form
   \be
   \label{eq:MF}
   \left(\frac{dn}{dM}\right)_{\rm NG}= \left(\frac{dn}{dM}\right)_{\mathrm{T},M_{300}}\left(\left.\frac{n_{\rm NG}}{n_\mathrm{G}}\right|_{{\rm Edgeworth}}\right)
    \ee
  where the first term on the right hand side is the Gaussian mass function of Tinker et al. \cite{Tinker:2008ff} for clusters identified as spheres containing a mean density 300 times that of the mean matter density of the Universe, $300\,\bar{\rho}_\mathrm{m}(z)$. The ratio of the non-Gaussian mass function to the Gaussian one will be given as a series expansion, defined below. This factor will be a function of mass, redshift, and parameters that characterize the amplitude of the non-Gaussianity, which we define next.

  \subsection{Parametrizing the level of non-Gaussianity}
  Since object number counts are not sensitive to the details of the momentum space correlations, we consider the dimensionless, connected moments (the cumulants, divided by the appropriate power of the amplitude of fluctuations) of the density fluctuations smoothed on a given scale $R$, as defined in Eq.(\ref{eq:Mns}). Most constraints on non-Gaussianity have so far been reported for a parameter that measures the size of the three-point correlation in momentum space, or bispectrum. This is an extremely useful first statistic because this correlation should be exactly zero if the fluctuations were exactly Gaussian. However, because the bispectrum is a function of two momenta, the non-Gaussian parameters most often quoted assume a shape for the bispectrum. 

  A generic homogeneous and isotropic bispectrum for the potential $\Phi$ can be written as
  \be
  \label{eq:threepoint}
  \Expect{\Phi(\vec{k}_1)\Phi(\vec{k}_2)\Phi(\vec{k}_3)}_c=(2\pi)^3\delta^3_D(\vec{k}_1+\vec{k}_2+\vec{k}_3)\;B(k_1,k_2,k_3)
  \ee
   where the function $B(k_1,k_2,k_3)$ determines the shape. Bispectra are colloquially named by the (triangle) configuration of the three momentum vectors that are most strongly correlated. To interpret our constraints on $\mathcal{M}_3$ in terms of familiar bispectra, we consider the templates for `local', `equilateral' and `orthogonal' bispectra:
  \ba
  \label{templates}
  B_{\rm local}&=&2f_{\rm NL}^{\rm local}(P(k_1)P(k_2) +P(k_1)P(k_3) +P(k_2)P(k_3))\\\nonumber
  B_{\rm equil}&=&6f_{\rm NL}^{\rm equil}[-P(k_1)P(k_2) +{\rm 2\;perm.} -2(P(k_1)P(k_2)P(k_3))^{2/3}\\\nonumber
  &&+P(k_1)^{1/3}P(k_2)^{2/3}P(k_3)+{\rm 5\;perm.} ]\\\nonumber
  B_{\rm orth}&=&6f_{\rm NL}^{\rm orth}[- 3P(k_1)P(k_2) +{\rm 2\;perm.}-8(P(k_1)P(k_2)P(k_3))^{2/3}\\\nonumber
  &&+3P(k_1)^{1/3}P(k_2)^{2/3}P(k_3) +{\rm 5\;perm.} ]
  \ea
  where the power spectrum, $P(k)$, is defined from the two-point correlation function by
  \be
  \Expect{\Phi(\vec{k}_1)\Phi(\vec{k}_2)}=(2\pi)^3\delta^3_D(\vec{k}_1+\vec{k}_2)P(k_1)\equiv(2\pi)^3\delta^3_D(\vec{k}_1+\vec{k}_2)2\pi^2\frac{\Delta^2_{\Phi}(k_0)}{k^3}\left(\frac{k_1}{k_0}\right)^{n_s-1}\;.
  \ee
  In the best fit cosmology from the seven-year WMAP data, baryon acoustic oscillations and Hubble parameter measurements, the spectral index is $n_s=0.967$ \cite{Komatsu:2010fb, Jarosik:2010iu}, and the amplitude is such that $\sigma_8=0.81$. Observationally, the parameter $f_{\rm NL}^{\rm local}$ is typically measured by looking for a bispectrum of the form given in the first line of Eq.(\ref{templates}), which has weaker implications than the definition of {\it all} the correlation functions as in Eq.(\ref{eq:localAnsatz}). 
  The best current limits on the amplitudes of the bispectra in Eq.(\ref{templates}) come from the {\it Planck} Satellite maps of the CMB \cite{Ade:2013tta}, which limit $f_{\rm NL}^{\rm local}=2.7\pm5.8$, $f_{\rm NL}^{\rm equil}=-42\pm75$, and $f_{\rm NL}^{\rm orth}=-25\pm39$ at the 68.3\% confidence level. Table \ref{table:M3} shows the value of $\mathcal{M}_3$, smoothed on a scale corresponding to $10^{14}\,h^{-1}\,\Msun$ halos ($h$ is related to the Hubble parameter today, $H_0=100 h \,{\rm km/s/Mpc}$), for the local, equilateral, and orthogonal templates using the WMAP7 best fit cosmology.
  \begin{table}[htbp]
  {\caption{The conversions between the parameter $\mathcal{M}_3$ and the amplitudes of particular bispectra $f_{\rm NL}$. These numbers assume the WMAP7 best fit cosmology and change by at most a few percent if the best fit cosmologies from our analysis are used instead. }\label{table:M3}}
    \begin{center}
   \begin{tabular}{|c|r@{\,}l|} 
    \hline
    \hline
    Shape & \multicolumn{2}{c|}{$\mathcal{M}_3$} \\ 
    \hline
    Local & $0.00031$&$f_{\rm NL}^{\rm local}$ \\
    Equilateral & $0.000086$&$f_{\rm NL}^{\rm equil}$\\
    Orthogonal & $-0.000062$&$f_{\rm NL}^{\rm orth}$\\
    \hline
    \end{tabular}
    \end{center}
  \end{table} 

  For some non-Gaussian scenarios (notably the local ansatz and typical single field models) the parameter $\mathcal{M}_3$ is interchangeable with the $f_{\rm NL}^{\rm local}$, $f_{\rm NL}^{\rm equil}$, or $f_{\rm NL}^{\rm orth}$ as a description of the amplitude of the three point function {\it and} as a measure of the total non-Gaussianity for the entire series of higher order correlations. This is possible when the cumulants scale parametrically as
  \be
  \Mh_n \propto \left(\mathcal{I}\mathcal{P}^{\frac{1}{2}}\right)^{n-2}\;,
  \label{eq:hierarchical}
  \ee
  where $\mathcal{I}$ is proportional to the appropriate $f_{\rm NL}$ parameter. Although it is not needed for the simplest models, we use this more general notation, since it is useful for the two-parameter scenarios we introduce below. The superscript `h' labels the scaling in Eq.(\ref{eq:hierarchical}), which we call {\it hierarchical}. 

  Mathematically, the higher order correlations could be nearly arbitrary, and it is only because we hope they have a common origin in some perturbation theory that it seems likely they are related. In this paper, we will contrast a second possible scaling, occurring in models where the scalar inflaton couples to a gauge field, that is much more non-Gaussian than the hierarchical scenario for a fixed value of the skewness \cite{Barnaby:2011pe, Barnaby:2012tk}. This scaling, which we call {\it feeder} (since it originates in models where fluctuations of a second field provide an extra source for the inflaton fluctuations\footnote{The equation of motion for the inflaton fluctuations $\delta\varphi$ is $[\partial_t^2-a^{-2}\nabla^2+3H\partial_t+m^2]\delta\varphi=J$, where $J$ is a source that depends on the quantum fluctuations of fields coupled to the inflaton \cite{Barnaby:2011pe}.}) is
  \be
  \Mf_n \propto \;\mathcal{I}^{\,n}\;,\;\;\;n\geq3.
  \ee
  Object number counts are sensitive to the value of the total skewness and to the scaling of higher moments, rather than any details of the momentum space correlations.

  In addition to the dependence on a parameter like $f_{\rm NL}$, the cumulants also have numerical coefficients that typically have to do with combinatorics. For example, beginning with Eq.(\ref{eq:localAnsatz}), the bispectrum contains three terms linear in $f_{\rm NL}^{\rm local}$, each with two equivalent ways to take the expectation value of pairs of fields $\Phi_\mathrm{G}$. We will the choose the constants of proportionality equal to combinatoric factors for the moments that are generated in the local ansatz and a simple two-field extension that gives feeder scaling:\footnote{The local ansatz was given in Eq.(\ref{eq:localAnsatz}) with $f^2_{\rm NL}\langle\Phi_\mathrm{G}(x)^{2}\rangle\ll 1$ to ensure weak non-Gaussianity. The moments generated have the hierarchical scaling with $\mathcal{I}=f_{\rm NL}$. To obtain representative combinatorics for the feeder scaling, we use a scenario where one Gaussian field and one subdominant but highly non-Gaussian field contribute to the inhomogeneities in the gravitational potential: $\Phi(x)=\phi_\mathrm{G}+\sigma_\mathrm{G}+\tilde{f}_{\rm NL}[\sigma_\mathrm{G}(x)^2-\langle\sigma_\mathrm{G}(x)^2\rangle]$, with $\tilde{f}_{\rm NL}\mathcal{P}_{\sigma}^{1/2}\gg 1$. In that case $\mathcal{I}=\tilde{f}_{\rm NL}\mathcal{P}_{\sigma}/\mathcal{P}_{\Phi}^{1/2}$.}
  \ba
  \label{eq:scalingh}
  \mathrm{Hierarchical}\hspace{5mm} &\Mh_n = n!\,2^{n-3}\left(\frac{\Mh_3}{6}\right)^{n-2}\\
  \label{eq:scalingf}
  \mathrm{Feeder}\hspace{5mm} &\Mf_n = (n-1)!\, 2^{n-1}\left(\frac{\Mf_3}{8}\right)^{n/3}.
  \ea 
  For a given scaling of the moments, we can determine a series expansion for the probability distribution and for the mass function that can be consistently truncated at some order in the moments.

  For the single parameter scenarios, we report constraints in terms of the scaling assumed and the parameter $\mathcal{M}_3$, which can be compared with other constraints on particular bispectrum shapes using Table \ref{table:M3}.

  \subsection{The mass function in terms of $\mathcal{M}_3$ and the scaling of higher moments} 
  We will assume the non-Gaussian factor in the mass function of Eq.(\ref{eq:MF}) takes the following form:
  \be
  \label{EdgeworthMassfcn}
  \left.\frac{n_\mathrm{NG}}{n_\mathrm{G}}\right|_{{\rm Edgeworth}}\approx1+\frac{F^{\rm h,f\prime}_1(M)}{F^{\prime}_0(M)}+\frac{F^{\rm h,f\prime}_2(M)}{F^{\prime}_0(M)}+\dots
  \ee
  Each term in the series is normalized by the Press-Schechter Gaussian term, $F^{\prime}_0(M)=(e^{-\nu_\mathrm{c}^{2}/2}/\sqrt{2\pi})( d\sigma/dM)(\nu_\mathrm{c}/\sigma)$, where $\nu_\mathrm{c}=\delta_\mathrm{c}/\sigma$, $\delta_\mathrm{c}=1.686$ is the collapse threshold, and $\sigma=\sigma(M)$ is the variance in density fluctuations smoothed on the appropriate scale (Eq.(\ref{eq:sigmaR})). Although the first term, $F^{\rm h\,\prime}_1(M)$ or $F^{\rm f\,\prime}_1(M)$, is proportional to $\mathcal{M}_3$ regardless of how the higher moments scale, the exact form of all higher order terms depends on the choice of scaling. For the hierarchical and feeder scaling, $F^{\rm h\,\prime}_n(M)$ and $F^{\rm f\,\prime}_n(M)$ are given in Eq.(\ref{eq:fprimes}) of the Appendix. Truncating this series after the first term is clearly unphysical since no probability distribution with only a non-zero skewness can be positive everywhere. Although for some objects (low mass, low redshift) this truncation does not cause a significant error, for rarer fluctuations it does. Keeping higher terms in the series is therefore important. How significant these terms are in the context of cluster constraints depends on the mass and redshift of the objects as well as the amplitude and scaling of the non-Gaussianity considered. In Section \ref{sec:OtherWork}, we show several examples to illustrate how relevant the higher terms are as a function of mass, redshift, skewness and scaling. Although this mass function has been shown to agree reasonably well with simulations, it does not come from a first principles derivation. In Section \ref{sec:OtherWork} we also contrast it to the Dalal et al mass function from simulations of the local ansatz \cite{Dalal:2007cu}.

  \section{Two parameter extensions}
  \label{sec:2param}
  We also consider the ability of the data to constrain four models characterized by two parameters, chosen to match classes of non-Gaussian inflation models. One very natural extension is to introduce an additional parameter, $0\leq q\leq1$, so that the hierarchical moments of the non-Gaussian density field behave as 
  \ba
  \label{eq:2LightFields}
  \mathcal{M}^{\rm h}_n&=&  \,n!\,2^{n-3}\,q\left(\frac{\mathcal{M}^{\rm h}_3}{6q}\right)^{n-2}\;.
  \ea
  Moments of this type arise in scenarios when two fields, one Gaussian and the other with weak local-type non-Gaussianity, contribute to the primordial fluctuations. In that case the total primordial gravitational perturbation is
  \be
  \label{eq:weak2field}
  \Phi(x)=\phi_\mathrm{G}(x)+\psi_\mathrm{G}(x)+\tilde{f}_{\rm NL}(\psi_\mathrm{G}^2-\langle\psi_\mathrm{G}^2\rangle)\;.
  \ee
  The ratio of the contribution of (the Gaussian part of) the field $\psi_\mathrm{G}$ to the total power is the parameter $q$:
  \be
  q\equiv\frac{\langle\psi_\mathrm{G}^2\rangle}{\langle\phi_\mathrm{G}^2\rangle+\langle\psi_\mathrm{G}^2\rangle}
  \ee
  and $\mathcal{I}=q\tilde{f}_{\rm NL}$. Since significant local non-Gaussianity only arises in multi-field models, this model is quite plausible. In the standard local ansatz, only one of the terms in Eq.(\ref{eq:weak2field}) contributes to the fluctuations in the gravitational potential, but that field is distinct from the field that sourced inflation. In other words, if $\phi_\mathrm{G}$ represents the fluctuations of the inflaton, the usual local ansatz corresponds to Eq.(\ref{eq:2LightFields}) with $q\rightarrow1$. In these inflation models, both fields must be very light compared to the Hubble scale during inflation, with masses $m_{1,2}\ll H$. 

  We can similarly introduce a second parameter into the feeder scaling by defining
  \ba
  \label{eq:2FieldsHeavyLight}
  \mathcal{M}^{\rm f}_n&=& q \,(n-1)! \, 2^{n-1}\left(\frac{\mathcal{M}^{\rm f}_3}{8q}\right)^{n/3}
  \ea
  again with $0\leq q\leq1$. Such a coefficient appears in the inflation scenario of \cite{Barnaby:2011pe} and in the `quasi-single field' models introduced by \cite{Chen:2009zp} where there is an additional heavy field relevant for the fluctuations (with mass very close to the Hubble scale during inflation). 

  For either choice of scaling for the cumulants, we will take $\mathcal{M}_3$ and $q$ as the free parameters. For fixed $\mathcal{M}_3$, smaller $q$ then corresponds to boosting the relative importance of higher order cumulants, while $q=1$ corresponds to our one-parameter models. For the scenarios to remain weakly non-Gaussian, we need $\mathcal{M}_3^{h,f}/q\ll1$. If our cosmology is consistent with Gaussian primordial fluctuations, we expect the data to favor $q \approx 1$ and small $\mathcal{M}_3$.

  As another extension of the one-parameter models, we add scale-dependence. One advantage of cluster number count constraints is the relatively small scale they probe compared to the CMB. This suggests that clusters may be particularly useful when used in conjunction with the CMB to constrain scale-dependent non-Gaussianity \cite{LoVerde:2007ri}. 
  To that end, we consider a scale-dependent ansatz with hierarchical scaling of the cumulants:
  \ba
  \label{eq:scaledep}
  \mathcal{M}_{3,R}&=&6\,\mathcal{I}_{R_0}\left(\frac{R}{R_0}\right)^{n_3},\\\nonumber
  \mathcal{M}_{n,R}&=&n!\,2^{n-3}\left(\frac{\mathcal{M}_3}{6}\right)^{n-2}\;,
  \ea 
  where $n_3$ is the additional model parameter. We chose the pivot scale to be $R_0 = 8h^{-1}$ comoving Mpc. Similarly, we can add scale dependence to the feeder type scenarios:
  \ba
  \label{eq:scaledepf}
  \mathcal{M}_{3,R}&=&8\,\,\mathcal{I}_{R_0}\left(\frac{R}{R_0}\right)^{n_3},\\\nonumber
  \mathcal{M}_{n,R}&=&(n-1)! \, 2^{n-1}\left(\frac{\mathcal{M}_3}{8}\right)^{n/3}\;.
  \ea

  \section{Application to galaxy cluster data}
  \label{sec:results}

  We now investigate the constraints on these models from current galaxy cluster data. In this work, we employ the data and analysis of \cite{Mantz:2009fw, Mantz:2009fx}, to which we refer the reader for complete details.

\begin{figure}[htb]
    \centering
   \includegraphics[width=0.45\textwidth]{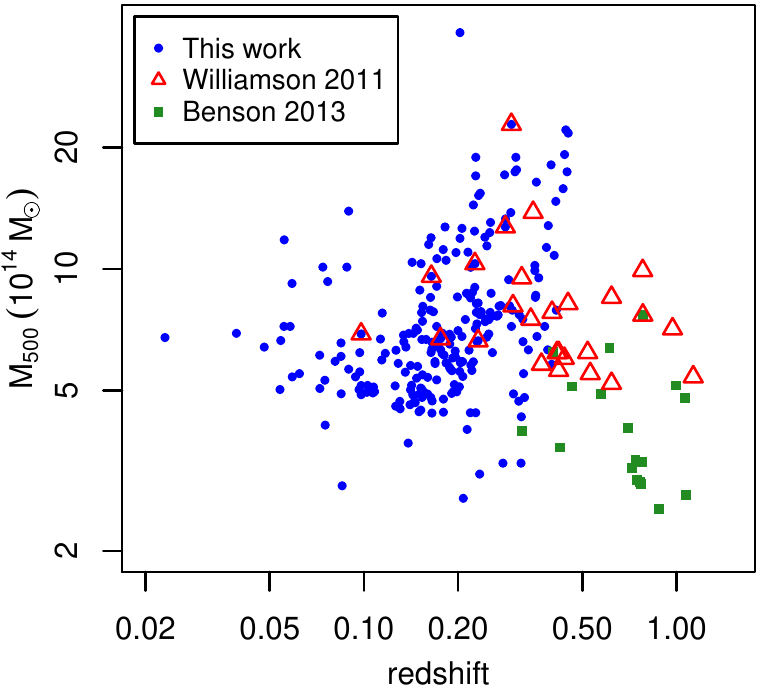}
    \caption{\label{fig:zM} For illustrative purposes, the mass and redshift distribution of our cluster data \cite{Mantz:2009fx} are compared with South Pole Telescope detected clusters used to obtain constraints on $f_{\rm NL}^{\rm local}$ by Williamson et~al. \cite{Williamson:2011jz} and Benson et~al. \cite{Benson:2011uta}. The mass values assume a flat concordance model with $h=0.7$ and $\Omegam=0.3$.}
  \end{figure}

  Briefly, the cluster data set consists of 237 X-ray bright clusters detected in the ROSAT All-Sky Survey \cite{Truemper:1993} and compiled in the BCS, REFLEX or bright MACS catalogs \cite{Ebeling:1998xm, Boehringer:2004fc, Ebeling:2010ra} satisfying conservative luminosity/flux thresholds such that the physical properties and selection function of the final sample are well understood.\footnote{Since the original work, Abell 689 has been removed from the data set, since {\it Chandra} observations reveal its X-ray emission to be dominated by a point source rather than the intracluster medium.} These detections span the redshift range $0<z<0.5$; in \figref\ref{fig:zM} their distribution in redshift and mass is compared to SZ detections that have been used in previous non-Gaussianity constraints. In addition to survey X-ray flux measurements and spectroscopic redshifts, archival ROSAT or {\it Chandra} data for 94 of the clusters were used to obtain mass proxies in the form of temperature and gas mass measurements. Total mass measurements from hydrostatic analysis of {\it Chandra} data for 42 relaxed clusters were also incorporated \cite{Allen:2007ue}.\footnote{Note that this sample of 42 only partly overlaps the larger cosmology sample, a fact which is accounted for in the data analysis.} Together these data provide the ability to calibrate the scaling and intrinsic scatter with mass of cluster X-ray luminosity, gas temperature and gas mass. This information is critical to accounting for survey selection effects and provides improved cosmological constraints generally compared with a pure self-calibration approach. Full details of our scaling relation model and the data used to constrain it are found in \cite{Mantz:2009fw, Mantz:2009fx}.

  Through a technique detailed in \cite{Allen:2007ue}, measurements of the ratio of gas mass to total mass for the 42 relaxed clusters above additionally and independently provide constraints on the mean cosmic matter density and the expansion of the Universe, which we also take advantage of here. We additionally incorporate seven-year WMAP constraints in some of the results below, using the publicly released WMAP data and analysis codes \cite{Jarosik:2010iu}. We do not model the effects of non-Gaussianity on the CMB power spectrum in this analysis, which is justified since even the current constraints limit any shift due to non-Gaussianity to be small; consequently, the WMAP data effectively only provide additional constraining power on the standard set of cosmological parameters, particularly $\sigma_8$.

Following the method outlined in \secref\ref{sec:mfcns}, we evaluate the non-Gaussian mass function as the product of a Gaussian mass function and a series whose terms depend on the model under consideration (see Eq.(\ref{eq:MF}) and Eq.(\ref{EdgeworthMassfcn})). As in \cite{Mantz:2009fw, Mantz:2009fx}, the Gaussian mass function we use is the simulation-calibrated fit of \cite{Tinker:2008ff}, with the cosmology-dependent correspondence of halo mass and $\sigma(M)$ provided by {\sc camb}.\footnote{\url{http://www.camb.info}} Note that, independent of any non-Gaussianity, our treatment includes an allowance for systematic uncertainty in the normalization, shape, and evolution of the Gaussian mass function in the form of a multidimensional prior on the relevant parameters \cite{Mantz:2009fw}. This prior translates to a 10\% uncertainty on the Gaussian mass function at the typical mass of our clusters ($10^{15}\Msun$). We evaluate the non-Gaussian series given by Eq.(\ref{EdgeworthMassfcn}) and Eq.(\ref{eq:fprimes}), keeping at most 16 terms for the hierarchical model and 17 for the feeder. This is a large enough number of terms that the series expansion of the PDF (estimated by the relative size of the first ignored term) is accurate to at least  20\% for the highest redshift, most massive clusters in the sample and $\mathcal{M}_3<0.04$ (0.025) for hierarchical (feeder) scaling. However, for the low values of $\mathcal{M}_3$ that the data prefers, a considerably smaller number of terms (around 5) is sufficient for the PDF to be accurate to a few percent in the range of interest, and there is not much difference in the mass function as the number of terms is changed. Furthermore, for larger levels of non-Gaussianity (outside the observational bounds, but still weakly non-Gaussian) the series can be badly behaved if too many terms are kept. See \cite{Shandera:2012ke} for more detail.

  Constraints on non-Gaussianity from clusters are limited by the precision with which cluster masses can be estimated, both individually (i.e. constraining the masses of the most rare objects) and statistically (constraining the mass function of the population). The former is straightforwardly related to the limitations of individual cluster measurements, while the latter also depends on the scaling and intrinsic scatter with mass of the survey observable used to define the cluster sample, as well as the ability of the data to constrain those nuisance parameters. In the present work, both our estimates of individual masses and the overall cluster mass scale include a systematic error budget of $\sim15\%$; these reflect allowances for a variety of uncertainties such as instrument calibration and departures from hydrostatic equilibrium \cite{Allen:2007ue, Mantz:2009fx}. Including all these allowances, our data ultimately provide a 10\% constraint on the normalization of the X-ray luminosity--mass relation, with the best fitting intrinsic scatter of that relation measured to be $(43\pm4)\%$ \cite{Mantz:2009fx}.

  When analyzing cluster data, we vary the mean baryon and matter densities (\Omegab{} and \Omegam{}), the Hubble parameter ($h$) and the amplitude of the matter power spectrum ($\sigma_8$), in addition to the parameters describing non-Gaussianity and cluster scaling relations, and a number of nuisance parameters accounting for various systematic uncertainties (see \cite{Allen:2007ue, Mantz:2009fw}). Priors on the Hubble parameter, $h=0.742 \pm 0.036$ \cite{Riess0905.0695}, and on the baryon density from Big Bang Nucleosynthesis (BBN), $\Omegab h^2 = 0.0214 \pm 0.002$ \cite{Kirkman:2003uv}, are included. When including WMAP data, we additionally marginalize over the optical depth to reionization, the spectral index of scalar fluctuations and its running with wavenumber, and the amplitude of small-scale CMB fluctuations due to the Sunyaev-Zel'dovich effect; the Hubble parameter and BBN priors are not used in this case. The full suite of our results are shown in Tables~\ref{tab:everything} and \ref{tab:convertFNLs}.

  \begin{figure}[tbp]
    \centering
    \includegraphics[width=0.45\textwidth]{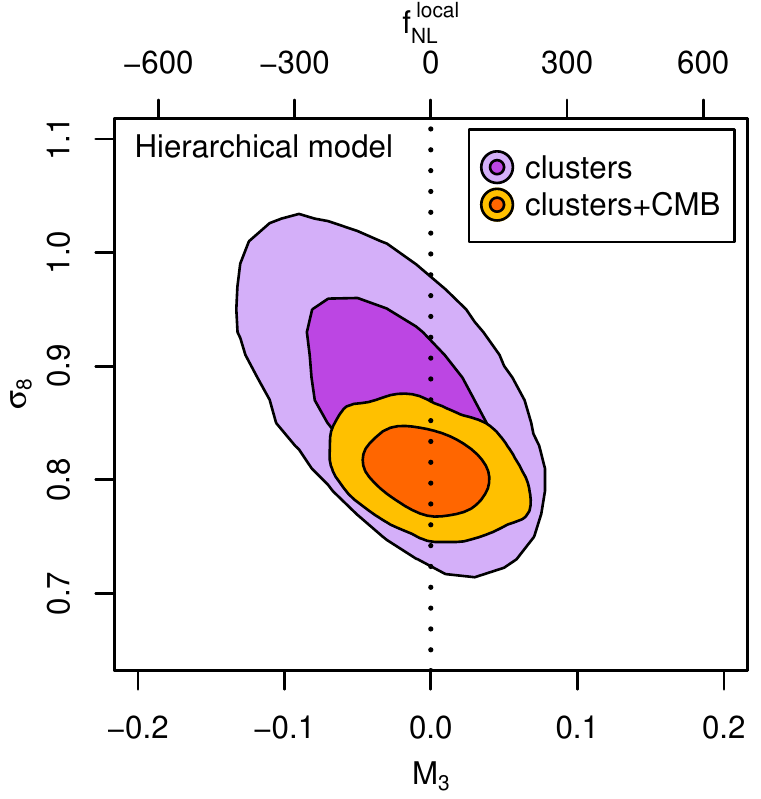}
    \hfill
    \includegraphics[width=0.45\textwidth]{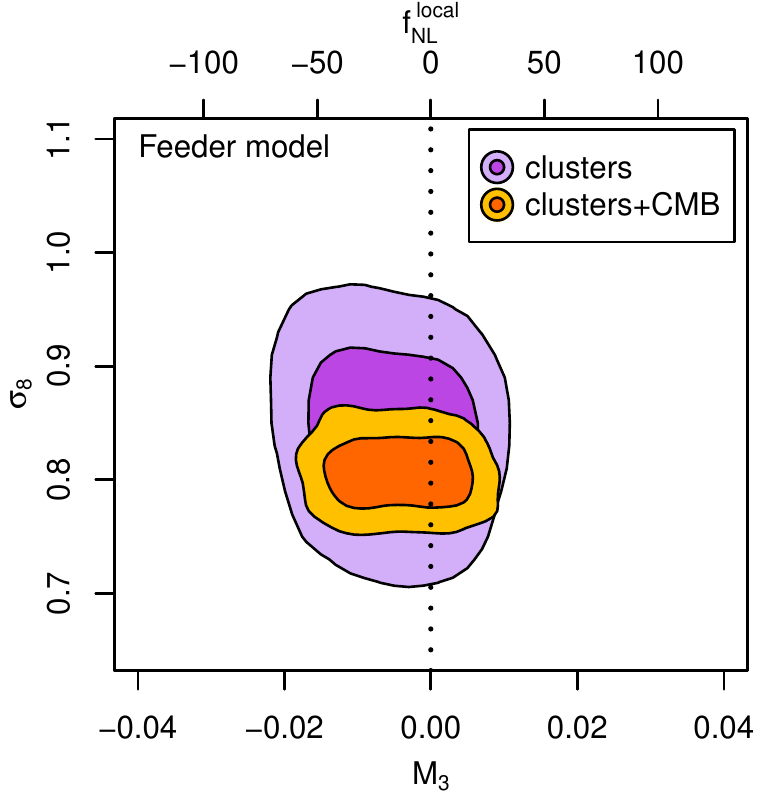}
    \caption{\label{fig:M3s8}Marginalized joint constraints on $\M_3$ (or the corresponding $f_{\rm NL}^{\rm local}$) and $\sigma_8$ at 68.3\% and 95.4\% confidence levels from cluster and clusters+CMB data, for single-parameter non-Gaussian models. Only CMB power spectra (not bispectra) are used in the combination, tightening constraints on the standard set of cosmological parameters, but providing no additional, direct constraining power on non-Gaussianity. The constraints are consistent with Gaussianity in all cases, but their strength and character depend on whether the hierarchical or feeder scaling (left and right panels) are used.}
  \end{figure}

  \begin{figure}[tbp]
    \centering
    \includegraphics[width=0.45\textwidth]{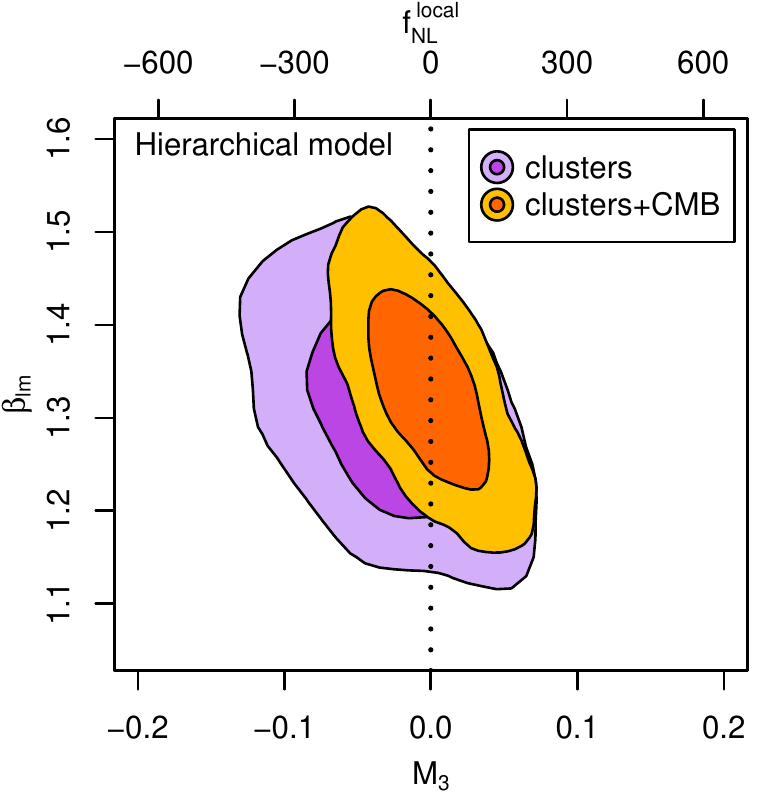}
    \hfill
    \includegraphics[width=0.45\textwidth]{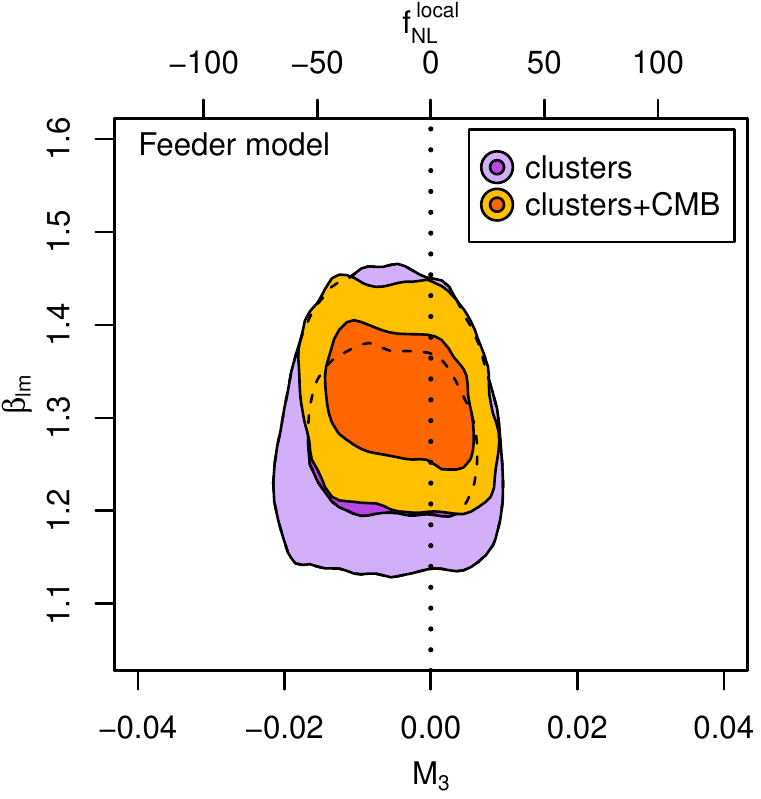}
    \caption{\label{fig:M3mlslope} As in \figref~\ref{fig:M3s8}, but showing constraints on $\M_3$ and $\beta_{\ell m}$, the power-law slope of the cluster X-ray luminosity--mass relation.}
  \end{figure}

  \renewcommand{\arraystretch}{1.25}
  \begin{table}[htbp]
   {\caption{Marginalized best fitting values and 68.3\% confidence intervals on the most interesting fit parameters (see text), using cluster (CL) and CL+CMB data, for various models (choices of the scaling and the number of free parameters). `skew-only' refers to truncating the series in Eq.(\ref{EdgeworthMassfcn}) after the second term, in which case there is no distinction between hierarchical (h) and feeder (f) scalings. For the scale-dependent models, the amplitude of the skewness is reported at the pivot point. $^a$A small local maximum in probability in the range $[-33,-28]$ is also formally included in this maximum likelihood confidence region (see also \figref~\ref{fig:lowz}).}\label{tab:everything}}
   \begin{center}
     \begin{tabular}{|c|c|r@{}l|c|c|}
       \hline\hline
       \multicolumn{1}{|c|}{Model} & Data & \multicolumn{2}{c|}{$10^3 \M_3$} & $\sigma_8$ & $\beta_{\ell m}$ \\
       \hline
       h & CL & $-23$&$^{+40}_{-35}$ & $0.86^{+0.07}_{-0.06}$ & $1.29^{+0.08}_{-0.07}$ \\
       h & CL $(z<0.3)$ & $113$&$^{+70}_{-65}$ & $0.76^{+0.08}_{-0.05}$ & $1.28^{+0.08}_{-0.10}$ \\
       h & CL+CMB & $-1$&$^{+24}_{-28}$ & $0.81^{+0.02}_{-0.03}$ & $1.33^{+0.07}_{-0.08}$ \\
       h+$q$ & CL+CMB & $-7$&$^{+28}_{-24}$ & $0.81^{+0.03}_{-0.02}$ & $1.34^{+0.07}_{-0.08}$ \\
       h+$n_3$ & CL+CMB & $3$&$^{+20}_{-30}$ & $0.81^{+0.02}_{-0.03}$ & $1.31^{+0.09}_{-0.06}$ \\
       f & CL & $-9$&$^{+11}_{-4}$ & $0.83^{+0.06}_{-0.04}$ & $1.27^{+0.07}_{-0.06}$ \\
       f & CL $(z<0.3)$ & $22$&$^{+11\,a}_{-19}$ & $0.84^{+0.06}_{-0.06}$ & $1.32^{+0.08}_{-0.09}$ \\
       f & CL+CMB & $-4$&$^{+7}_{-7}$ & $0.81^{+0.02}_{-0.02}$ & $1.32^{+0.06}_{-0.05}$ \\
       f+$q$ & CL+CMB & $-1$&$^{+4}_{-5}$ & $0.81^{+0.02}_{-0.02}$ & $1.32^{+0.06}_{-0.05}$ \\
       f+$n_3$ & CL+CMB & $-5$&$^{+8}_{-6}$ & $0.80^{+0.02}_{-0.02}$ & $1.32^{+0.07}_{-0.05}$ \\
       \hline
       skew-only & CL & $-9$&$^{+164}_{-24}$ & $0.81^{+0.05}_{-0.09}$ & $1.26^{+0.07}_{-0.06}$ \\ 
       skew-only & CL+CMB & $-3$&$^{+72}_{-20}$ & $0.80^{+0.03}_{-0.03}$ & $1.28^{+0.08}_{-0.04}$ \\
       \hline
     \end{tabular}
   \end{center}
  \end{table}
  \renewcommand{\arraystretch}{1.0}

  \renewcommand{\arraystretch}{1.25}
  \begin{table}[htbp]
   \caption{The constraints on the skewness can be converted to constraints on the amplitude of any bispectrum. The shape of the bispectrum is independent of the scaling, although the usual local ansatz corresponds to a local-shape bispectrum with hierarchical moments. \label{tab:convertFNLs}}
   \begin{center}
    \begin{tabular}{|c|c|r@{}l|r@{}l|r@{}l|} 
    \hline
    \hline
    Scaling & Data & \multicolumn{2}{c|}{Local Bispectrum} & \multicolumn{2}{c|}{Equil. Bispectrum} & \multicolumn{2}{c|}{Orthog. Bispectrum} \\
   \hline
   h & CL & \hspace{7mm}$-73$&$^{+129}_{-113}$ & \hspace{6.5mm}$-271$&$^{+482}_{-422}$ & \hspace{11mm}$346$&$^{+538}_{-615}$ \\
   h & CL+CMB & $-3$&$^{+78}_{-91}$ & $-12$&$^{+289}_{-338}$ & $15$&$^{+430}_{-369}$ \\
   f & CL &   $-28$&$^{+35}_{-13}$ & $-106$&$^{+134}_{-48}$ & $130$&$^{+60}_{-164}$\\
   f & CL+CMB & $-14$&$^{+22}_{-21}$ & $-52$&$^{+85}_{-79}$ & $63$&$^{+97}_{-104}$ \\
  \hline
   skew-only & CL & $-29$&$^{+532}_{-78}$ & $-105$&$^{+1916}_{-280}$ & $146$&$^{+389}_{-2658}$\\ 
   skew-only & CL+CMB & $-9$&$^{+234}_{-65}$ & $-35$&$^{+841}_{-234}$ & $48$&$^{+324}_{-1167}$ \\
   \hline
   \end{tabular}
    \end{center}
  \end{table}

  \renewcommand{\arraystretch}{1.0}

  Marginalized joint constraints on $\M_3$ and $\sigma_8$ are shown in \figref\ref{fig:M3s8} for the single-parameter non-Gaussian model, using both hierarchical and feeder scalings. In both cases, the principal degeneracy of $\M_3$ is with $\sigma_8$ when only cluster data are used. This is intuitive, since $\sigma_8$ determines the relative rarity of massive clusters even in the purely Gaussian case. With the addition of CMB data, $\sigma_8$ is independently tightly constrained, and for both scalings the primary degeneracy of $\M_3$ is instead with the slope of the cluster X-ray luminosity--mass relation, $\beta_{\ell m}$, as shown in \figref\ref{fig:M3mlslope}. This reflects the dependence of the number of \emph{detections} of massive clusters in an observable-limited survey on the relevant scaling relation. Compared with the hierarchical scaling, the degeneracy between $\M_3$ and other model parameters is generically weaker in the feeder model. This is due to the relatively larger high-order non-Gaussian moments generated by the feeder scaling, whose effect is less easily mimicked by changing other parameters (Figure~\ref{fig:fNL30}; see also Figure~1 of \cite{Barnaby:2012tk}).

\begin{figure}[tbp]
    \centering
    \includegraphics[width=0.45\textwidth]{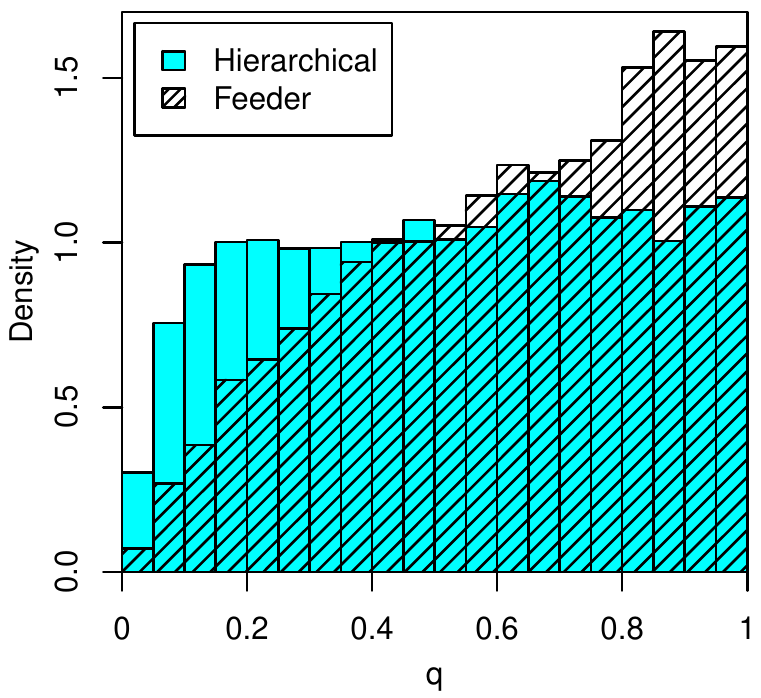}
    \caption{\label{fig:qhist} Posterior probability density of the $q$ parameter for hierarchical and feeder scalings. In our parametrization, small values of $q$ boost higher order non-Gaussian moments relative to $\M_3$ and $q=1$ corresponds to the one-parameter models with either scaling. The clusters+CMB data disfavor very small values of $q$, consistent with the overall preference for Gaussian initial fluctuations, although only the feeder scaling shows a clear (if modest) preference for $q=1$. At 95.4\% confidence, the lower limits on $q$ are respectively $q>0.10$ and $q>0.18$ for the hierarchical and feeder scenarios.}
  \end{figure}

  We also consider two-parameter non-Gaussian models for the two scalings, including either the $q$ or $n_3$ parameters introduced in Section \ref{sec:2param}. In neither case is the additional parameter well constrained by the data, although the impacts on the $\M_3$ constraints are relatively minor (Table~\ref{tab:everything}). However, we find that values  $q \approx 0$, corresponding to having strongly non-Gaussian fluctuations regardless of the amplitude of $\M_3$, are disfavored (see \figref\ref{fig:qhist}). The clusters+CMB data  respectively provide 95.4\% confidence lower limits of $q>0.10$ and $q>0.18$ with the hierarchical and feeder scalings, assuming a uniform prior from zero to one.

  For ease of comparison to the literature (Section~\ref{sec:OtherWork}), we also obtained constraints on $\mathcal{M}_3$ keeping only the first term in the non-Gaussian mass function, proportional to the skewness. As shown in Table \ref{tab:everything}, the resulting error bars are larger than when we include more terms in the series, reflecting the fact that those terms generically increase the deviation of the mass function from the Gaussian one at higher masses and higher redshifts (again, see Figure \ref{fig:fNL30}). For example, keeping only the first term, our constraints from cluster and CMB data correspond to $f^{\rm local}_{\rm NL}=-9^{+234}_{-78}$ (Table~\ref{tab:convertFNLs}). When including all relevant terms we find $f^{\rm local}_{\rm NL}=-3^{+78}_{-91}$ for the hierarchical scaling and $f^{\rm local}_{\rm NL}=-14^{+22}_{-21}$ for the feeder scaling.

\begin{figure}[tbp]
    \centering
    \includegraphics[width=0.45\textwidth]{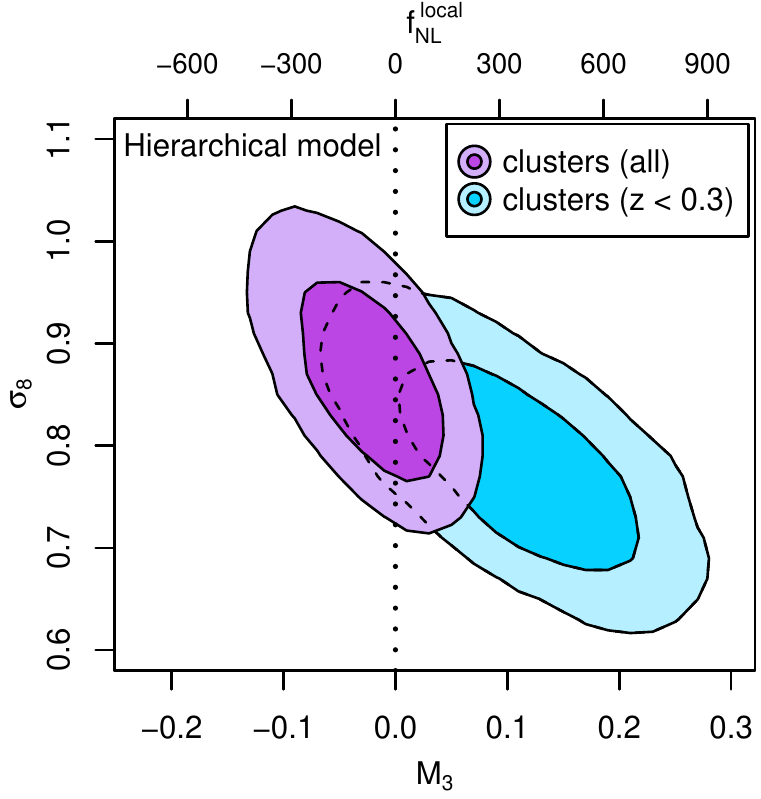}
    \hfill
    \includegraphics[width=0.45\textwidth]{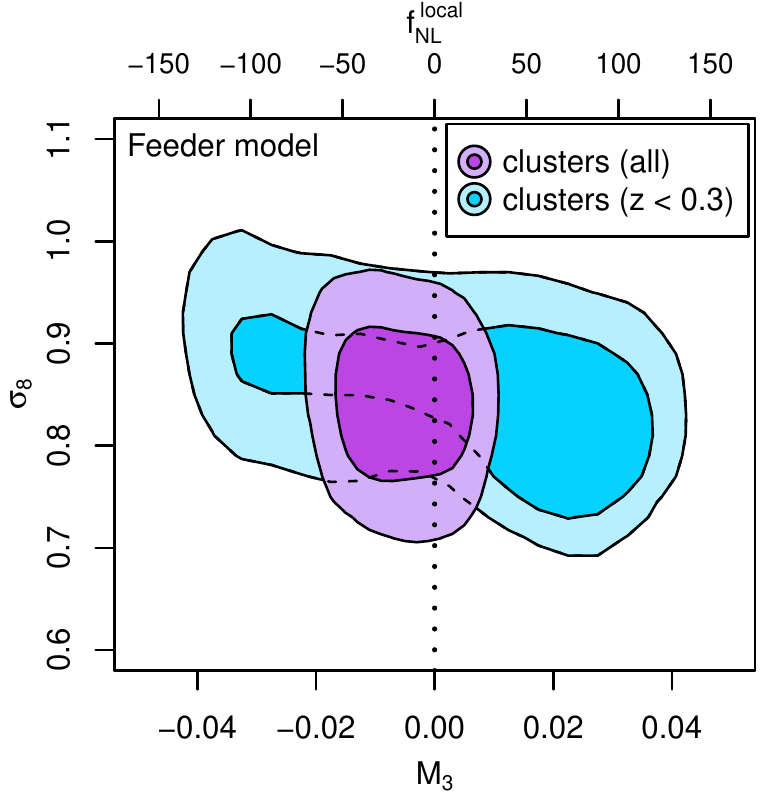}
    \caption{\label{fig:lowz} As in \figref~\ref{fig:M3s8}, but comparing constraints obtained from the full cluster data set with those from only clusters at redshifts $z<0.3$.}
  \end{figure}

  In light of the favorable comparison of our results to those obtained previously (see Section~\ref{sec:OtherWork}), we briefly investigate what characteristics of our data set influence the results. There is a practical limitation, however, since any attempt to reduce the overall size of the sample, the number of clusters with mass estimates from follow-up data, or the mass/redshift ranges covered, necessarily impacts constraints on the full set of cosmological and scaling relation parameters. Consequently, we confine ourselves to a single, limited, but informative comparison by asking how our constraints change when data at $z \geq 0.3$ are excluded. In detail, this low-redshift sample contains 203 clusters, of which 61 have follow-up data, compared to 237 and 94 for the full data set. As shown in \figref~\ref{fig:lowz} for single-parameter non-Gaussian models using the full hierarchical and feeder scalings, the constraining power of this low-redshift data set is significantly reduced. Results for the low-redshift clusters only are shown in Table~\ref{tab:everything}.

  \section{Comparison with the literature}
  \label{sec:OtherWork}
  Previous forecasts for constraints on non-Gaussianity from cluster counts have been done by Pillepich et al. \cite{Pillepich:2011zz} for the eROSITA X-ray mission, Sartoris et al. \cite{Sartoris:2010cr} for future X-ray surveys resembling the Wide Field X-ray Telescope concept, Oguri \cite{Oguri:2009ui} for a variety of future optical surveys, Cunha et al. \cite{Cunha:2010zz} for optically selected clusters in the Dark Energy Survey (DES), and Mak and Pierpaoli \cite{Mak:2012yb} for future surveys using the Sunyaev-Zel'dovich effect. There have been three previous cluster constraints on non-Gaussianity: two based on clusters detected in the SPT survey, by Benson et al. \cite{Benson:2011uta}, who find $f_{\rm NL}^{\rm local}=-192\pm310$, and Williamson et al. \cite{Williamson:2011jz}, who report $f_{\rm NL}^{\rm local}=20\pm 450$; and one based on the SDSS maxBCG cluster catalogue, by Mana et al. \cite{Mana:2013qba}, who have $f_{\rm NL}^{\rm local}=282\pm317$.\footnote{This result corresponds to their analysis of only cluster number counts, without including either the cluster power spectrum or CMB data.} 

  The existing forecasts and constraints use a variety of prescriptions for the non-Gaussian mass function (listed in Table \ref{table:MFs}). These mass functions are given in Eq.(\ref{eq:fs}) in the Appendix and are plotted in Figure \ref{fig:fNL30}. The levels of non-Gaussianity shown are $\mathcal{M}_3=0.009$ and $\mathcal{M}_3=0.031$, which correspond to the local model bispectrum with $f^{\rm local}_{\rm NL}=30$ and $100$, respectively. Notice that, below about $10^{15}$ \Msun, the Dalal et al. mass function \cite{Dalal:2007cu} deviates a little less from the Gaussian for a given value of $f_{\rm NL}$ than the LoVerde et al. mass function \cite{LoVerde:2007ri} (hereafter LMSV). However, some authors have found the LMSV mass function agrees better with simulation results if a reduced collapse threshold, $\delta_c\sim1.5$, is used. If that adjustment is made, the Dalal et al mass function would deviate more from the Gaussian than LMSV; see \cite{Pillepich:2008ka} for a comparison of all these cases. Since the Dalal et al. mass function was calibrated on simulations of the local ansatz, in principle it should include information about higher moments. This technique, though, has only been tried against one set of simulations and only for non-Gaussianity of the local type. A more precisely calibrated, more general non-Gaussian mass function will be important for any future analysis of non-Gaussianity with clusters.

  \begin{table}[htbp]
  {\caption{Gaussian mass functions and non-Gaussian extensions used in the literature. The non-Gaussian mass functions are either the first order semi-analytic expression from LoVerde et al. \cite{LoVerde:2007ri} (LMSV) or the mass function calibrated on N-body simulations of the local ansatz by Dalal et al. \cite{Dalal:2007cu}. All non-Gaussian mass functions also make use of a Gaussian mass function such as those fit by Sheth and Tormen \cite{Sheth:1999su}, Warren et al. \cite{Warren:2005ey}, Jenkins et al. \cite{Jenkins:2000bv} or Tinker et al. \cite{Tinker:2008ff, Tinker:2010}.}\label{table:MFs}}
    \begin{center}
    \begin{tabular}{|c|c|} 
    \hline
    \hline
     Author & Mass Function used \\ 
    \hline
    Benson \cite{Benson:2011uta}&  Jenkins + $f_{\rm Dalal}$ \\
    Cunha \cite{Cunha:2010zz}&   Jenkins + $f_{\rm Dalal}$ \\
    Mak \cite{Mak:2012yb}& Tinker + $f_{{\rm LMSV, skew\,only} }$  \\
    Mana \cite{Mana:2013qba}& Tinker + $f_{{\rm LMSV, skew\,only} }$  \\
    Oguri \cite{Oguri:2009ui} & Warren +$f_{{\rm LMSV, skew\,only}}$  \\
    Pillepich \cite{Pillepich:2011zz} & Tinker + $f_{{\rm LMSV, skew\,only}}$ \\
    Sartoris \cite{Sartoris:2010cr}&  Sheth-Tormen + $f_{\rm LMSV, skew\,only}$ \\
    Williamson \cite{Williamson:2011jz} &  Jenkins + $f_{\rm Dalal}$ \\
  \hline
  This work & Tinker + $f_{{\rm LMSV, many\, terms}}$ \\
    \hline
    \end{tabular}
    \end{center}
  \end{table}

  \begin{figure}[tbp]
  \begin{center}
  $\begin{array}{cc}
  \includegraphics[width=0.5\textwidth,angle=0]{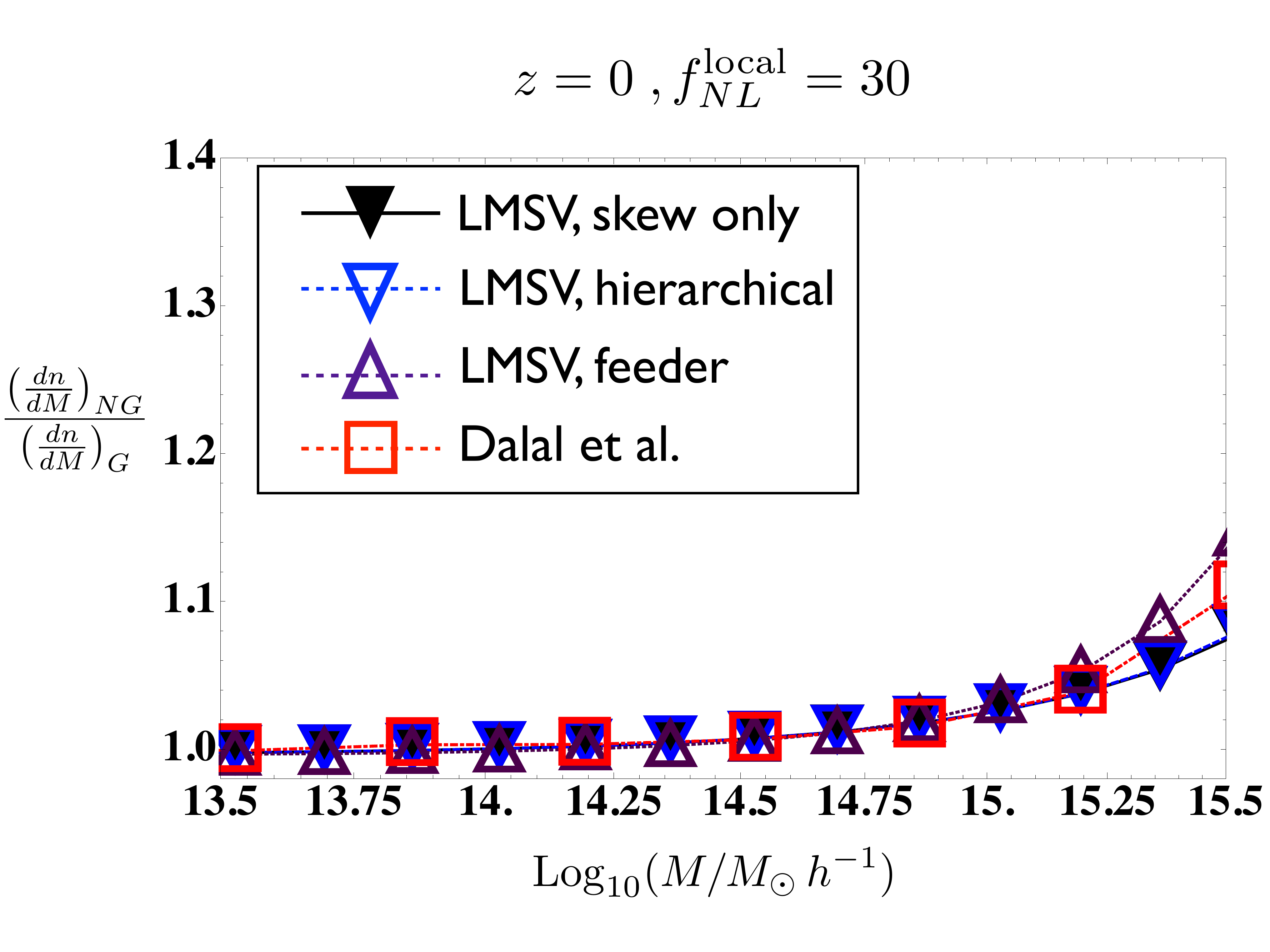} &
  \includegraphics[width=0.5\textwidth,angle=0]{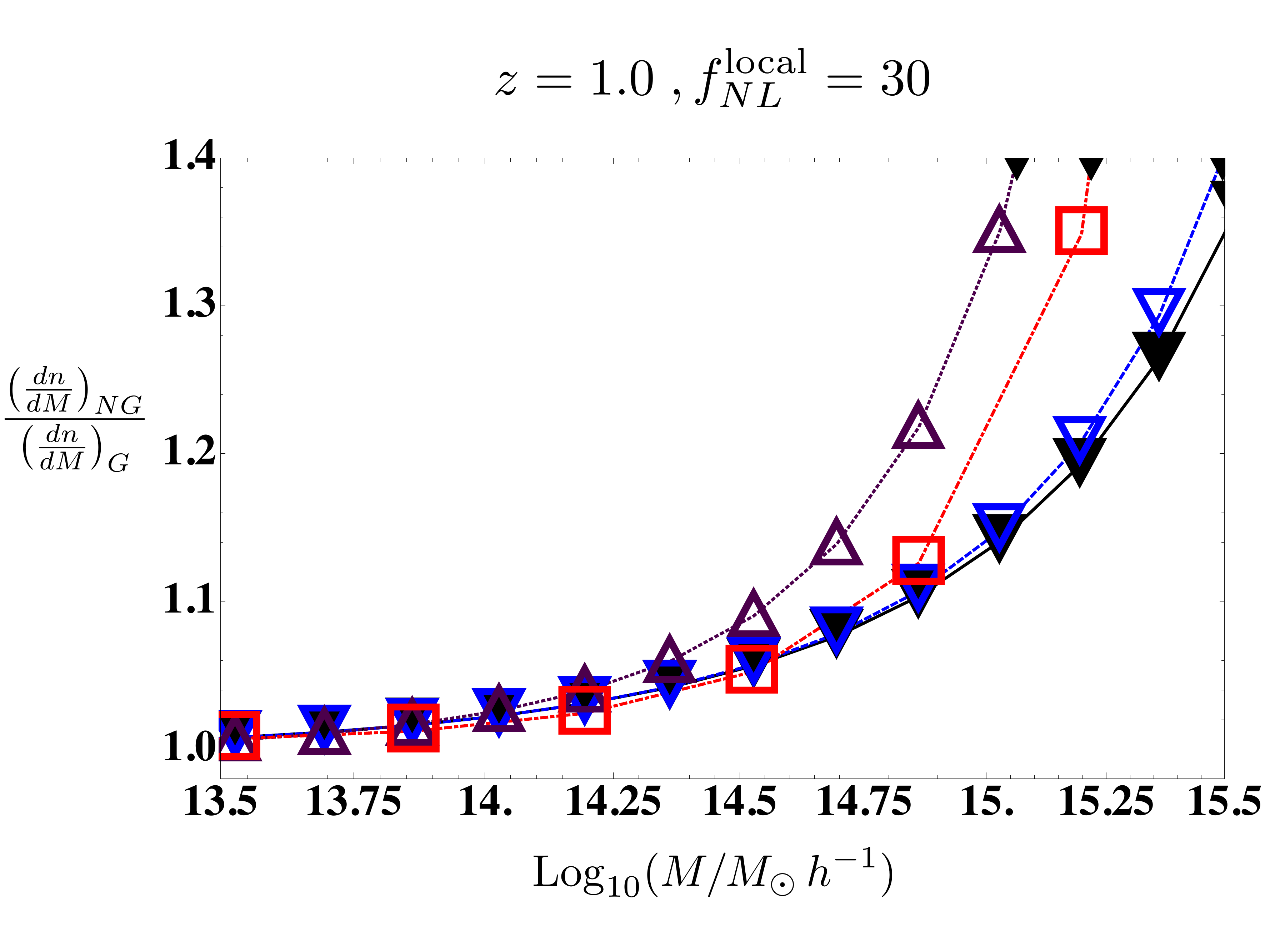} \\
  \includegraphics[width=0.5\textwidth,angle=0]{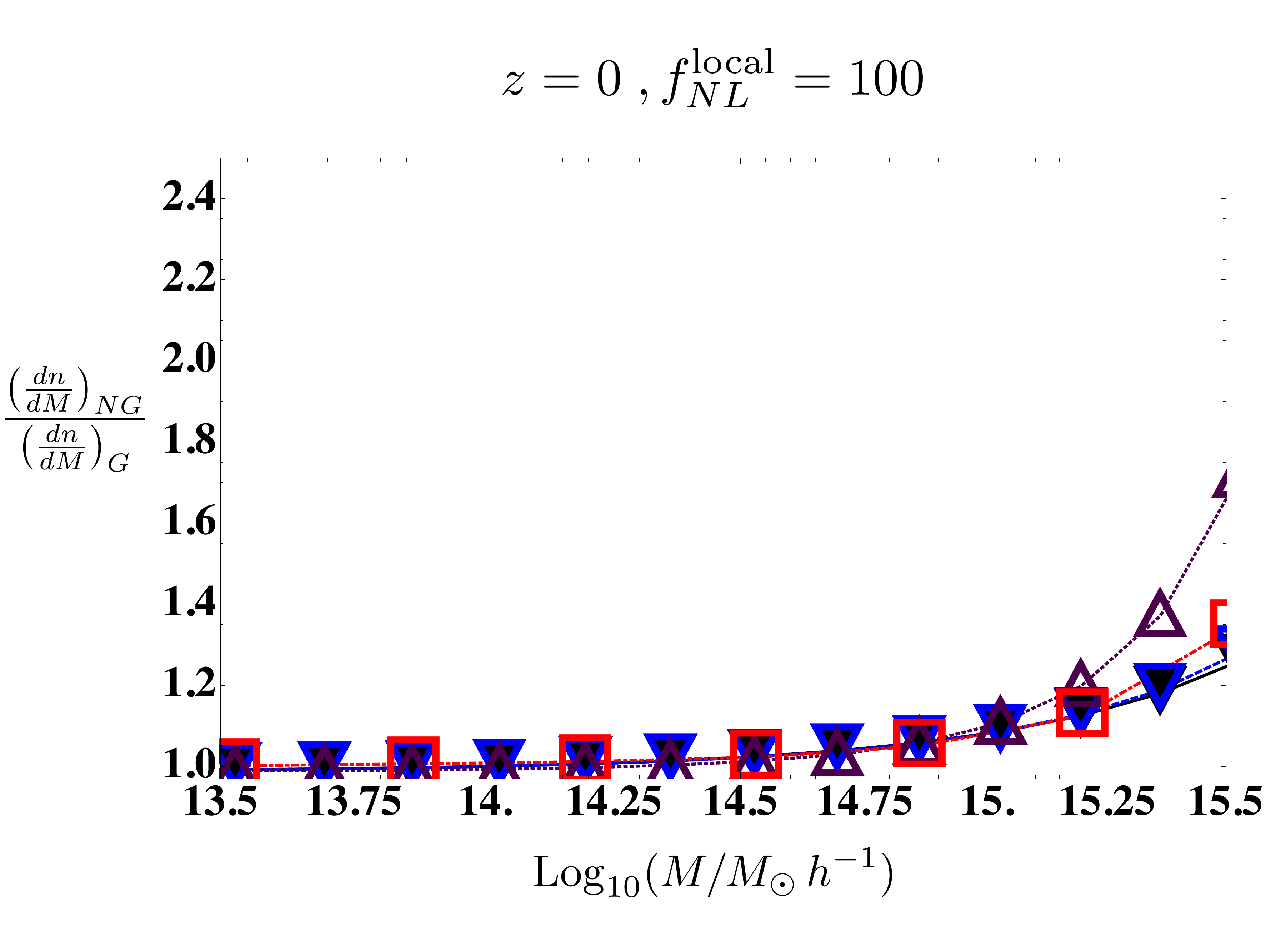} &
  \includegraphics[width=0.5\textwidth,angle=0]{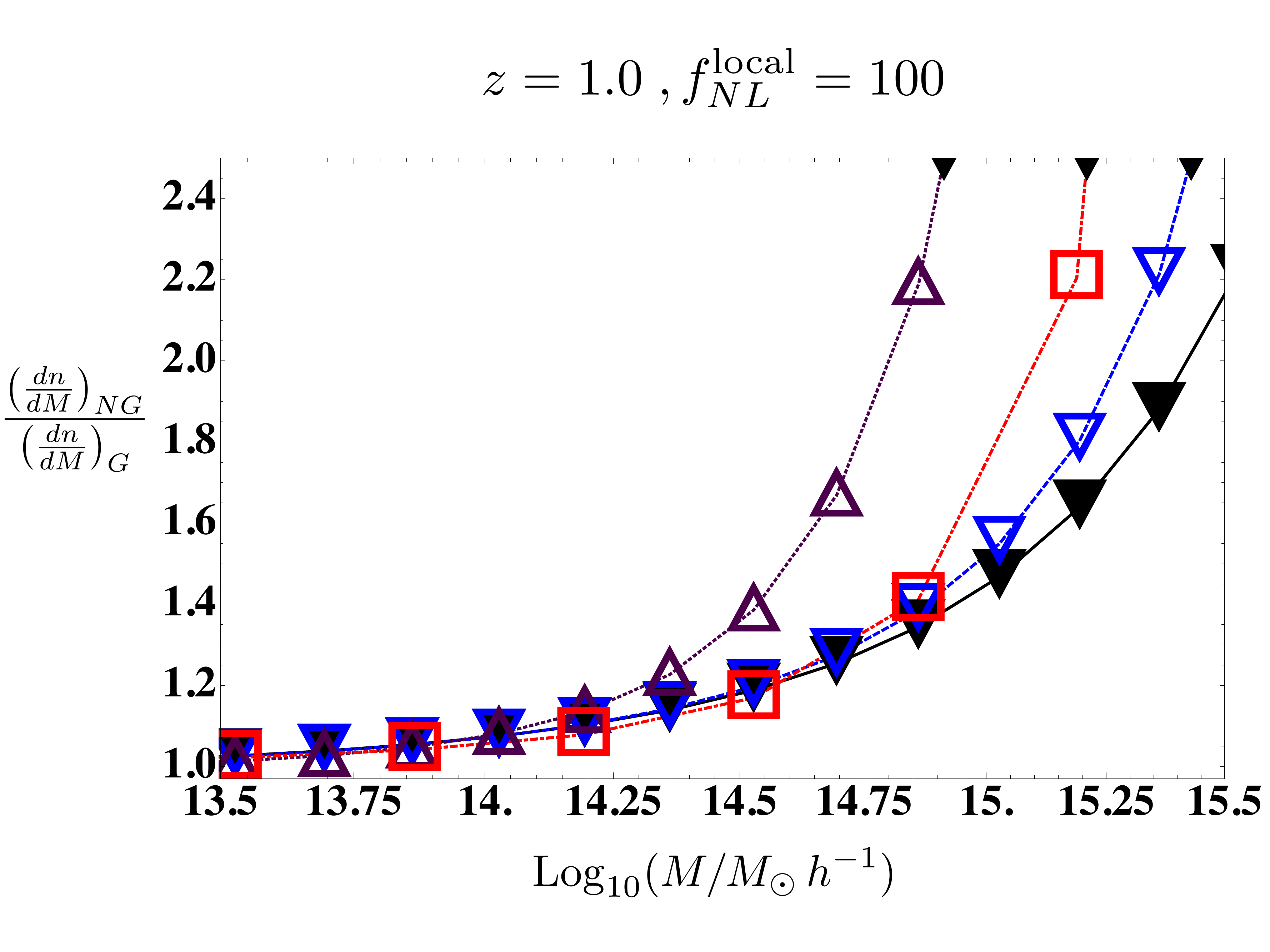}\\
  \end{array}$
  \caption{Comparing mass functions. Black (filled point down triangles): LMSV with one term. Blue (open point down triangles): LMSV with hierarchical scaling, keeping 4 terms. Purple (open point up triangles): LMSV with feeder scaling, keeping seven terms. (The number of terms was chosen to give good behavior up to $f_{\rm NL}=300$ at redshift 1.) Red (squares): Dalal et al. + Tinker et al. Note the difference in scale on the vertical axes in the top row compared to the bottom row.\label{fig:fNL30}}
  \end{center}
  \end{figure}

  Apart from the non-Gaussian mass function, these forecasts and analyses differ from one another and from ours in two principal ways: the form and complexity assumed for the mass--observable relation and its intrinsic scatter, and priors on the associated parameters. The most pessimistic forecasts in the literature find marginalized one sigma errors on $f_{\rm NL}^{\rm local}$ around $\mathcal{O}(10^3)$ (e.g, some cases analyzed in \cite{Cunha:2010zz,Pillepich:2011zz}). Those results assume that the scaling relations will be constrained solely through self-calibration \cite{Majumdar:2003mw} rather than with estimates of cluster masses, which can significantly boost the constraining power \cite{Wu:2009wg}. In addition, some forecasts assume significant photometric redshift errors \cite{Pillepich:2011zz}. As outlined in Section \ref{sec:results}, all of the clusters in our sample have spectroscopic redshifts and for nearly half we also have follow-up X-ray data that significantly improve the mass determinations. 

  Among the SPT results, Benson et~al. use a smaller area of the survey than Williamson et~al., but have an improved mass calibration and extend their sample to lower SZ detection significance (i.e. lower mass). In comparison, our cluster data set is significantly larger than either of the SPT cluster samples, contains more massive clusters (although at lower redshifts), has a larger intrinsic scatter in the mass--observable relation (although the parameters of the scaling relation are better constrained), and uses a more straightforward mass calibration (i.e. directly incorporating X-ray mass measurements rather than calibrating mass via simulation priors or priors based on external X-ray data). The masses and redshifts of these three cluster samples are compared in Figure~\ref{fig:zM}.

  Both of the SPT studies used CMB data to constrain the amplitude of the Gaussian power spectrum, as well as other cosmological parameters, as we did in Section~\ref{sec:results}. Both also used the Dalal et~al. non-Gaussian mass function, so it is not immediately clear which of our constraints is most comparable to theirs. However, comparing the various mass functions for the masses and redshifts of the SPT clusters and non-Gaussianity of magnitude $f_{\rm NL}^{\rm local}\sim\mathcal{O}(100)$, the better comparison seems to be with our skew-only results (Figure~\ref{fig:fNL30}). Our clusters+CMB constraint for the skew-only model corresponds to $f^{\rm local}_{\rm NL}=-9^{+234}_{-65}$, roughly a factor of two tighter than the constraints of Benson et~al., but less of an improvement than a straightforward dependence on the cluster sample size (237 versus 18) would imply. The precision of the mass calibration used in the two works is similar, and most likely limits the improvement we see from the larger data set. The greater constraining power of the Benson et~al. analysis (18 clusters) versus that of Williamson et~al. (26 clusters) further underscores the impact of providing additional data to improve the cluster mass calibration. On the other hand, the significant improvement in the constraining power of our full data set compared with only the low-redshift portion (see Figure~\ref{fig:lowz}) indicates that our current constraints are still not entirely systematically limited.

  Compared to both our data and the SPT samples, the maxBCG catalogue used by Mana et~al.  \cite{Mana:2013qba} contains a very large number (13\,823) of mostly less massive clusters at low redshifts ($0.1<z<0.3$) \cite{Koester:2007}. Their mass calibration is accomplished through a stacked weak gravitational lensing analysis, which constrains the mean optical richness--mass relation but does not provide mass estimates for individual clusters \cite{Sheldon:2009, Johnston:2007}. This feature, as well as the lack of high mass and high redshift clusters, presumably limits their constraining power. We note, however, that the large range in mass probed allows them to achieve from cluster counts \emph{alone} constraints comparable to those of Benson et~al. from clusters+CMB. Our skew-only, clusters-only constraint ($f^{\rm local}_{\rm NL}=-29^{+532}_{-78}$) is comparable to that of Mana et~al.

 There are some existing constraints on primordial non-Gaussianity beyond the skewness that are complementary to those we find here. For example, the CMB data from the WMAP satellite has been analyzed by \cite{Fergusson:2010gn, Ade:2013tta}. The {\it Planck} data has not yet been exhaustively analyzed for evidence of a trispectrum (a four-point correlation function), but the collaboration has reported the strongest bound to date for one of the simplest momentum-space trispectrum shapes. This shape is part of the local family of non-Gaussianity (it is the four point function generated by Eq.(\ref{eq:localAnsatz})), but its amplitude can be constrained independently of the amplitude of the three-point function ($f_{\rm NL}^{\rm local}$) and is typically called $\tau_{\rm NL}$. {\it Planck} data constrains the amplitude of this shape to be $\tau_{NL}<2800$ \cite{Ade:2013tta}, which is much weaker than the constraint implied by {\it Planck}'s tight limits on $f_{NL}^{\rm local}$ for a single parameter model with hierarchical scaling. However, the weakness of this constraint is likely to be due in large part to the remaining foreground and systematic effects present in and yet to be removed from the current analysis. For comparison, our central value in the hierarchical case, $\mathcal{M}_3=-0.001$ corresponds to a model with $\tau_{\rm NL}\approx 9$ (using the fitting function of \cite{LoVerde:2011iz} at $10^{14}h^{-1}\Msun$). Our central value for the feeder scaling, which is much more non-Gaussian, corresponds to $\tau_{\rm NL}\approx 12\,300$. By introducing the parameter $q$, we also tested models where the relationship between the skewness and the kurtosis was relaxed, although the rest of the series was still determined in terms of the two parameters. The constraints on non-Gaussianity from {\it Planck} data using Minkowski functionals do incorporate information from the series of cumulants in a way similar to cluster number counts. These could be more directly compared to our results with some additional analysis. 
  
  Recently Giannantonio et al \cite{Giannantonio:2013uqa} reported a constraint on non-Gaussianity from a clustering analysis of a broad ensemble of multiwavlength galaxy surveys. In that analysis, $f_{NL}^{\rm local}$ is degenerate with a trispectrum parameter$g_{NL}$ (the amplitude of the bispectrum generated by adding a term proportional to $\Phi_G^3$ to the local ansatz). For $f_{NL}^{\rm local}=0$ they report $-4.5\times 10^5<g_{NL}<1.6\times10^5$ (95\% C.L.) while if $f_{NL}^{\rm local}$ is nonzero and positive they find that $g_{NL}$ can take somewhat more negative values. Our results, using the hierarchical scaling appropriate for weakly non-Gaussian local models, are comparable. Again using a fitting formula from \cite{LoVerde:2011iz} at $10^{14}h^{-1}\Msun$, our largest allowed magnitude for the kurtosis ($|\mathcal{M}_4|\sim5\times10^{-3}$ for the one-parameter model with approximate 95\% bounds $|\mathcal{M}_3|\lesssim60\times 10^{-3}$) corresponds to $|g_{NL}|\lesssim 9\times10^4$. Allowing the two-parameter model with $q\sim0.1$ relaxes this constraint to $|g_{NL}|\lesssim 9\times10^5$. If the same trispectrum shape is considered with feeder scaling for the rest of the moments, our results imply $|g_{NL}|\lesssim 3\times10^5$ for the single parameter model and $|g_{NL}|\lesssim 6\times10^5$ with $q=0.18$.

  \section{Conclusions}
  \label{sec:Conclude}
  We have used the mass and redshift distribution of X-ray bright clusters to find that the primordial inhomogeneities in the gravitational potential can be consistently described by a Gaussian distribution to an accuracy of about one part in $10^3$ at scales around $0.1$--$0.5\, h\;{\rm Mpc}^{-1}$. Our constraints apply to any model of weak non-Gaussianity that has a sufficiently regular ordering in the cumulants to be modeled by one of the expressions in Section \ref{sec:mfcns} or Section \ref{sec:2param} (Equations (\ref{eq:scalingh}), (\ref{eq:scalingf}), (\ref{eq:2LightFields}), (\ref{eq:2FieldsHeavyLight}), (\ref{eq:scaledep}) or (\ref{eq:scaledepf})). In particular, for non-Gaussianity described by the local ansatz (Eq.(\ref{eq:localAnsatz}), which defines all cumulants for the model), we find $f_{\rm NL}^{\rm local}=-3^{+78}_{-91}$ from cluster data combined with the WMAP 7-year data. 

  Our analysis differs from previous work in that we include higher order cumulants in the non-Gaussian mass function. This allows us to differentiate constraints on two different one-parameter models for the non-Gaussianity, characterized by the relevance of moments beyond the skewness. We also tested the sensitivity to several two-parameter models, but found that current data are not very sensitive to this extra level of detail. Our full set of results can be found in Table \ref{tab:everything}. Table \ref{tab:convertFNLs} shows those results interpreted in terms of several popular models for the bispectrum. 

  The {\it Planck} satellite data recently led to very tight constraints on any non-Gaussianity on a significant range of scales, but those bounds are still about two orders of magnitude above the minimal levels predicted by slow-roll inflation. Some combination of lower-redshift probes will be required to explore the rest of that parameter space. This is an extremely worthwhile pursuit, since higher order correlations (or their absence) is our only route to learn more about the primordial era. The previously unexplored sensitivity of cluster data to higher order moments of non-Gaussianity shown by our results argues for revisiting the forecasts for future cluster surveys. Further bounds on non-Gaussianity remain as theoretically interesting as more precise measurements of the dark energy equation of state. Since the next stage of research on the primordial era will be dominated by large scale structure observations, it is crucial to understand the full potential of complementary information from cluster number counts together with CMB and large scale structure constraints on the bispectrum and trispectrum, and the halo bias.

  There are several ways our analysis could be extended. First, in order to consistently treat as large a family of non-Gaussian models as possible, we have used a simple, semi-analytic form for the non-Gaussian mass function. It would be preferable to use non-Gaussian mass functions calibrated on simulations, but those results do not yet exist for a sufficient variety of scenarios. However, simulations of a two-field local model that spans between the hierarchical scenario (with the extra parameter $q$) and the feeder behavior are in progress \cite{Adhikari:2013}. Once this simulation work is complete, we could revisit our analysis using the Dalal et~al. mass function for the local ansatz together with a comparable expression for models whose higher moments are relatively more important.

  The cluster data sets that have been used to constrain non-Gaussianity so far differ qualitatively in several respects. Nevertheless, empirical comparison of their constraining power underscores that the precision of the overall mass calibration, the availability of mass estimates for individual clusters, and the mass and redshift ranges probed all have an important role. In the very near term, combining the available survey data as well as gravitational lensing data (e.g. \cite{vonderLinden:2012kh, Applegate:2012kr}) in a multi-wavelength analysis has significant potential.

  A variety of upcoming survey data could also improve on the non-Gaussianity constraints in complementary ways. Data from, for example, the optical-wavelength Dark Energy Survey \cite{Abbott:2005bi} and the upcoming  eROSITA all-sky X-ray survey \cite{Merloni:2012uf} will extend the mass range of the cluster samples downward over a wide redshift range, and enable clustering analyses. Continued SZ surveys will be crucial for detecting the most massive, high redshift clusters which are most sensitive to non-Gaussianity. In addition, targeted X-ray follow-up of individual clusters will allow the more precise mass measurements that significantly aid the overall statistical power of cluster samples. All of this data is being collected to study important outstanding problems in cosmology, especially dark energy and neutrino mass, but is also well suited to further tests of non-Gaussianity to study inflation. Constraints on the primordial fluctuations from cluster counts and clustering provide a complementary cross-check to the CMB and to other large scale structure probes, and will continue to be an important tool for cosmology.

  \acknowledgments
  S.S. thanks Adrienne Erickcek for carefully checking the properties of higher order terms in the Petrov expansion. We also thank Neal Dalal, Dragan Huterer, Marilena LoVerde and Christian Reichardt for useful comments on a draft of this article. S. Shandera is supported by the Eberly Research Funds of The Pennsylvania State University. The Institute for Gravitation and the Cosmos is supported by the Eberly College of Science and the Office of the Senior Vice President for Research at the Pennsylvania State University. A. Mantz is supported by NSF grant AST-0838187. D. Rapetti acknowledges support from the DARK Fellowship program at the Dark Cosmology Centre, which is funded by the Danish National Research Foundation. Calculations in this work used the Coma and Orange compute clusters at the SLAC National Accelerator Laboratory. This work was supported in part by the U.S. Department of Energy under contract number DE-AC02-76SF00515.

  \appendix
  \section{Non-Gaussian mass functions}
  \label{appendix:MF}
  To model the effect on non-Gaussianity on the cluster mass function, we use an extended Press-Schechter approach \cite{Press:1973iz}, following previous work in \cite{Chiu:1997xb, Robinson:1999se, Matarrese:2000iz, LoVerde:2007ri, LoVerde:2011iz}. 
  The non-Gaussian probability distribution of (normalized) density fluctuations smoothed on a scale $R$ (associated to halos of mass $M$ by $R=\left((3/4\pi\rho)M\right)^{1/3}$) is $P(\nu,M)$. Here $\nu=\delta/\sigma_R$ (with $\sigma_R$ being the smoothed variance). The fraction of volume in collapsed objects (halos) is
  \be
  \label{eq:collapseFraction}
  F(M)=2\int_{\delta_\mathrm{c}/\sigma_R}^\infty\!\!d\nu P(\nu,M)
  \ee
  with $\delta_\mathrm{c}$ the threshold for collapse, and where we have included the Press-Schechter factor of 2 in front of the integral.  Then the number density $dn(M)/dM$ of halos with masses between $(M,M+dM)$ is
  \be
  \frac{dn}{dM}(M,z)=-2\frac{\bar{\rho}}{M}\frac{dF}{dM}\;,
  \label{eq:dndmdef}
  \ee
  where $\bar{\rho}=\Omegam\rho_\mathrm{crit}$ is the average (comoving) matter density. The smoothed density field is given by
  \be
  \delta_R(z)=\int \frac{d^3k}{(2\pi)^3} W_R(k)\delta(\vec{k},z)
  \ee
  where $W_R(k)$ is the Fourier transform of a window function, which we take to be a top-hat in real space.
  The smoothed variance is 
  \be
  \sigma^2(M,z)=\frac{1}{2\pi^2}\int dk\, P_{\rm lin}(k,z)W^2(k,M) k^2\;.
  \label{eq:sigmaR}
  \ee
  The matter perturbations $\delta$
  at redshift $z$ are related to the perturbations in the early matter era
  potential $\Phi$ by
  \begin{eqnarray}
  \delta(\vec{k},z)&=&M(k,z)\Phi(\vec{k})\\[0.2cm]\nonumber
  M(k,z)&=&\frac{2}{3}\frac{1}{\Omegam}\frac{c^2}{H_0^2}D(z)\frac{g(0)}{g(\infty)}T(k)k^2\;.
  \end{eqnarray}
  Here $\Omegam$ is the matter
  density relative to critical, $H_0$ is the Hubble constant, $D(z)$ is the
  linear growth function at redshift $z$ normalized to one today, and the
  growth suppression factor is $g(z=0)/g(z=\infty)\simeq0.76$ in the
  best-fit $\Lambda$CDM model. The variance of density fluctuations at 
  redshift $z$ smoothed on a scale $R$ associated to mass $M$ is 
  $\sigma^2(M,z)$, defined by
  \begin{equation}
  \sigma^2(M, z)=\int_0^{\infty}\frac{dk}{k}W_R(k)^2M(k,z)^2\Delta^2_{\Phi}(k)\;,
  \label{eqn:sigma}
  \end{equation}
  where $\Delta^2_{\Phi}(k)$ is the amplitude of fluctuations, related to the power spectrum by $P_{\Phi}(k)=(2\pi^2/k^3)\Delta^2_{\Phi}(k)\left(k/k_0\right)^{n_s-1}$.

  Petrov \cite{Petrov:1972, Petrov:1987} (references are the English translations) developed an asymptotic expansion for non-Gaussian PDFs, which is a generalization of the Edgeworth expansion. In terms of the dimensionless smoothed moments $\mathcal{M}_{n,R}$, this is
  \ba
  \label{eq:PetrovM}
  P(\nu){\rm d}\nu&=&\frac{d\nu}{\sqrt{2\pi}}e^{-\nu^2/2}\left\{1+\sum_{s=1}^{\infty}\sum_{\{k_m\}}{\rm H}_{s+2r}(\nu)\prod_{m=1}^2\frac{1}{k_m!}\left(\frac{\mathcal{M}_{m+2,R}}{(m+2)!}\right)^{k_m}\right\}\;.
  \ea
  Here $H_n(\nu)$ are Hermite polynomials defined by $H_n(\nu)=(-1)^ne^{\nu^2/2}\frac{{\rm d}^n}{{\rm d}\nu^n}e^{-\nu^2/2}$ and $r=k_1+k_2+\dots+k_n$ where the set $\{k_m\}$ is built of all non-negative integer solutions of
  \be
  \label{eq:Diophantine}
  k_1+2k_2+\dots+nk_n=s\;.
  \ee
  We use the two scalings, hierarchical and feeder, from Eq.(\ref{eq:scalingh}) or Eq.(\ref{eq:scalingf}), to organize the terms in this series. Then we can write 
  \be
  F(M)=F_0(M)+F_1(M)+F_2(M)+\dots
  \ee
  where 
  \be
  F_0(M)=\frac{1}{2}{\rm Erfc}\left(\frac{\nu_\mathrm{c}}{\sqrt{2}}\right)
  \ee
  and the rest of the series is ordered according to the scaling:
  \ba
  \label{eq:Fexpand}
  F^{\rm h}(M)&=&\frac{1}{2}{\rm erfc}\left(\frac{\nu_{\mathrm{c}}}{\sqrt{2}}\right)+\frac{e^{-\nu_{\mathrm{c}}^2/2}}{\sqrt{2\pi}}\sum_{s=1}^{\infty}\sum_{\{k_m\}_h}{ H}_{s+2r-1}(\nu_{\mathrm{c}})\prod_{m=1}^s\frac{1}{k_m!}\left(\frac{\mathcal{M}_{m+2,R}}{(m+2)!}\right)^{k_m}\\\nonumber
  F^{\rm f}(M)&=&\frac{1}{2}{\rm erfc}\left(\frac{\nu_{\mathrm{c}}}{\sqrt{2}}\right)+\frac{e^{-\nu_{\mathrm{c}}^2/2}}{\sqrt{2\pi}}\sum_{s=1}^{\infty}\sum_{\{k_m\}_f}{H}_{s+1}(\nu_{\mathrm{c}})\prod_{m=1}^s\frac{1}{k_m!}\left(\frac{\mathcal{M}_{m+2,R}}{(m+2)!}\right)^{k_m}\;.
  \ea
  The sets $\{k_m\}_h$ are again non-negative integer solutions to $k_1+2k_2+\dots+sk_s=s$ and $r=k_1+k_2+\dots+k_m$, but the sets $\{k_m\}_f$ are solutions to $3k_1+4k_2+\dots+(s+2)k_s=s+2$. Now for either scaling, truncating the series at some finite $s$ in the sums above keeps all terms up to the same order in $\mathcal{M}_3$: $\mathcal{M}^s_3$ for hierarchical scalings and $\mathcal{M}^{s/3}_3$ for feeder scalings.

  To write the mass function we will need derivatives of all the terms in the expansion with respect to mass (or smoothing scale). In general, the derivatives can be found using the relationship for the Hermite polynomials:
  \be
  \nu H_n(\nu)-\frac{dH_n(\nu)}{d\nu}=H_{n+1}(\nu)\;.
  \ee
  The ratio of the non-Gaussian Edgeworth mass function to the Gaussian has the same structural form for either scaling:
  \be
  \label{EdgeworthMassfcn}
  \left.\frac{n_\mathrm{NG}}{n_\mathrm{G}}\right|_{{\rm Edgeworth}}\approx1+\frac{F^{\rm h,f\prime}_1(M)}{F^{\prime}_0(M)}+\frac{F^{\rm h,f\prime}_2(M)}{F^{\prime}_0(M)}+\dots
  \ee
  with the derivatives of each term $F_s^{\prime}=dF_s/dM$ for $s\geq1$:
  \ba
  \label{eq:fprimes}
  F_s^{\rm h\,\prime}(\nu)&=&F_0^{\prime}\sum_{\{k_m\}_h}\left\{H_{s+2r}\prod_{m=1}^s\frac{1}{k_m!}\left(\frac{\mathcal{M}_{m+2,R}}{(m+2)!}\right)^{k_m}\right.\\\nonumber
  &&\left.+H_{s+2r-1}\frac{\sigma}{\nu}\frac{d}{d\sigma}\left[\prod_{m=1}^s\frac{1}{k_m!}\left(\frac{\mathcal{M}_{m+2,R}}{(m+2)!}\right)^{k_m}\right]\right\}\\\nonumber
  F_s^{\rm f\,\prime}(\nu)&=&F_0^{\prime}\sum_{\{k_m\}_f}\left\{H_{s+2}\prod_{m=1}^s\frac{1}{k_m!}\left(\frac{\mathcal{M}_{m+2,R}}{(m+2)!}\right)^{k_m}\right.\\\nonumber
  &&\left.+H_{s+1}\frac{\sigma}{\nu}\frac{d}{d\sigma}\left[\prod_{m=1}^s\frac{1}{k_m!}\left(\frac{\mathcal{M}_{m+2,R}}{(m+2)!}\right)^{k_m}\right]\right\}\\\nonumber
  \ea
  where the $\{k_m\}$ again satisfy the relationships given below Eq.(\ref{eq:Fexpand}) and we have used
   \be
  F_0^{\prime}=\frac{e^{-\frac{\nu_\mathrm{c}^2}{2}}}{\sqrt{2\pi}}\frac{d\sigma}{dM}\;\frac{\nu_\mathrm{c}}{\sigma}\;.
  \ee
We have shared computer code to evaluate series for these two scalings at \url{http://www.slac.stanford.edu/~amantz/work/nongauss13/}.

  \subsection{Other non-Gaussian mass functions}

  The mass function above has been shown to agree relatively well with $N$-body simulations of the local ansatz \cite{Pillepich:2008ka, LoVerde:2011iz}, although there is some disagreement in the literature about just how well it agrees \cite{Kang:2007gs}. Some work has found evidence of a better fit if one shifts the collapse threshold \cite{Grossi:2009an}, but our data are not sensitive to that level of detail. Ideally, one would like to use mass functions explicitly calibrated on simulations. To that end, Dalal et al. \cite{Dalal:2007cu} proposed the following
  \be
  \frac{dn}{dM}=\int dM_\mathrm{G}\frac{dn}{dM_\mathrm{G}}\frac{dP}{dM}(M_\mathrm{G})
  \ee
  where $dn/dM_\mathrm{G}$ is the Gaussian mass function (which Dalal et al. took to be Jenkins et al. \cite{Jenkins:2000bv}) and $dP/dM(M_\mathrm{G})$ is the probability distribution describing how a Gaussian halo of mass $M_\mathrm{G}$ maps into a non-Gaussian halo with mass $M$. Dalal et al. gave fitting formula for this as a Gaussian distribution with:
  \ba
  \left\langle\frac{M}{M_\mathrm{G}}\right\rangle-1 &=& 1.3\times 10^{-4} f_{\rm NL}\sigma_8\sigma(M_\mathrm{G},z)^{-2}\\\nonumber
  {\rm var}\left(\frac{M}{M_\mathrm{G}}\right)&=&1.4\times 10^{-4}(f_{\rm NL}\sigma_8)^{0.8}\sigma(M_\mathrm{G},z)^{-1}\;.
  \ea
  For a fixed value of $f_{\rm NL}$, this mass function deviates less from the Gaussian than the LMSV mass function does. This is an interesting approach, but has not been tried for non-local non-Gaussianity. 

  For reference, we collect here the various non-Gaussian mass functions that exist in the literature. The mass function can be parametrized as
  \be
  \frac{dn}{dM}(M,z)=f(\sigma)\frac{\bar{\rho}}{M}\frac{d\ln[\sigma^{-1}(M,z)]}{dM}\;,
  \label{eq:universalMF}
  \ee
  where $f(\sigma)$ will differ between various models. For reference, we compile here $f(\sigma)$ for the (Gaussian) Tinker mass function \cite{Tinker:2008ff} ($f_{\mathrm{T},\Delta=300}$), the LMSV mass function \cite{LoVerde:2007ri} truncated at first order ($f_{{\rm LMSV}, 1}$), the Dalal et al. \cite{Dalal:2007cu} mass function ($f_{\rm Dalal}$), and the LMSV mass function keeping a large number ($n$) of terms with either hierarchical ($f_{{\rm LMSV},n}^{\rm h}$) or feeder ($f_{{\rm LMSV},n}^{\rm f}$) scaling for the cumulants:
  \ba
  \label{eq:fs}
  f_{\mathrm{T},\Delta=300}(\sigma)&=&A\left[\left(\frac{\sigma}{b}\right)^{-a}+1\right]e^{-c/\sigma^2}\;,\\\nonumber
  f_{{\rm LMSV}, 1}&=&f_{\mathrm{T},\Delta=300}\left[1+\frac{F^{\rm h,f\prime}_1(M)}{F^{\prime}_0(M)}\right]\;,\\\nonumber
  &=&f_{\mathrm{T},\Delta=300}\left[1+H_3\frac{\mathcal{M}_3}{3!}+H_2\frac{\sigma}{\nu}\frac{d\mathcal{M}_3}{d\sigma}\right]\;,\\\nonumber
  f_{\rm Dalal}(\sigma_M)&=&\left[\frac{d\ln[\sigma^{-1}(M,z)]}{dM}\right]^{-1}\int dM_\mathrm{G}\,f_{\mathrm{T},\Delta=300}(\sigma_{M_\mathrm{G}})\frac{d\ln[\sigma^{-1}(M_\mathrm{G},z)]}{dM_\mathrm{G}}\\\nonumber
  &&\times\frac{1}{\sqrt{2\pi}\sigma_{M_\mathrm{G}}}{\rm exp}\left[-\frac{(\frac{M}{M_\mathrm{G}}-\langle\frac{M}{M_\mathrm{G}}\rangle)^2}{2\sigma_{M/M_\mathrm{G}}^2}\right]\;,\\\nonumber
  f_{{\rm LMSV},n}^{\rm h,f}&=&f_{\mathrm{T},\Delta=300}\left[1+\sum_{s=1}^{n}\frac{F_s^{\rm h,f\prime}}{F_0^{\prime}}\right]\;.
  \ea
  In general, anywhere we have written $f_{\mathrm{T},\Delta=300}$ in the non-Gaussian mass functions, one can substitute any other Gaussian model. For models with hierarchical scaling, it is common to use the skewness parameter $S_3=\mathcal{M}_3/\sigma$. With that definition, it is clear that the second line is equivalent to the expression used for the LMSV mass function in \cite{Pillepich:2011zz,Mak:2012yb},
   \be
   f_{{\rm LMSV}, 1}=f_{\mathrm{T},\Delta=300}\left[1+\frac{1}{6}\frac{\sigma^2}{\delta_\mathrm{c}}\left[S_3\left(\frac{\delta_\mathrm{c}^4}{\sigma^4}-2\frac{\delta_\mathrm{c}^2}{\sigma^2}-1\right)+\frac{1}{6}\frac{dS_3}{d\ln\sigma}\left(\frac{\delta_\mathrm{c}^2}{\sigma^2}-1\right)\right]\right]\;.
   \ee

\providecommand{\href}[2]{#2}\begingroup\raggedright\endgroup

\end{document}